\documentclass[journal]{IEEEtran}

\usepackage{cite}
\usepackage{graphicx}    
\usepackage{amsmath,amssymb,amsfonts, bm}
\usepackage{algorithm,algcompatible}
                         
\usepackage{textcomp}      
\usepackage{changepage}   % for the adjustwidth environment
\usepackage{balance}
\usepackage[export]{adjustbox}

\usepackage{caption}
\usepackage{subcaption}
\usepackage{color}

\usepackage{algorithm,algcompatible}
\usepackage{booktabs}

\newtheorem{lemma}{\bf Lemma}
\newtheorem{assumption}{\bf Assumption}
\newtheorem{theorem}{\bf Theorem}
\newtheorem{remark}{\bf Remark}

\newenvironment{proof}{{\it {Proof:}}}{\hfill$\blacksquare$}

\DeclareMathOperator{\bR}{\mathbb R}
\DeclareMathOperator{\BR}{\mathbb R}
\DeclareMathOperator{\bE}{\mathbb E}
\DeclareMathOperator{\col}{\mathrm{col}}
\DeclareMathOperator{\diag}{\mathrm{diag}}

\DeclareMathOperator{\tp}{{\mathsf{T}}}

\renewcommand{\baselinestretch}{0.95}

\begin{document}

	\title{
Dual Control of Exploration and Exploitation for Auto-Optimisation Control with Active Learning
}

	\author{Zhongguo Li, \IEEEmembership{Member, IEEE},
			Wen-Hua Chen, \IEEEmembership{Fellow, IEEE}\\
		Jun Yang, \IEEEmembership{Fellow, IEEE},
		 Yunda Yan, \IEEEmembership{Member, IEEE}
	\thanks{This work was supported by the UK Engineering and Physical Sciences Research Council (EPSRC) Established Career Fellowship ``Goal-Oriented Control Systems: Disturbance, Uncertainty and Constraints" under the grant number EP/T005734/1. Corresponding author: Wen-Hua Chen.}
	\thanks{Z. Li is with Department of Electrical and Electronic Engineering, University of Manchester, Manchester, M13 9PL U.K. (email: zhongguo.li@manchester.ac.uk).}
		\thanks{W.-H. Chen and J. Yang are with Department of Aeronautical and Automotive Engineering, Loughborough University, Loughborough,  LE11 3TU, U.K. (emails: w.chen@lboro.ac.uk;  j.yang3@lboro.ac.uk). }
			\thanks{Y. Yan is with Department of Computer Science, University College London, London, WC1E 6BT, U.K. (email: yunda.yan@ucl.ac.uk).}
	}

\maketitle

{\color{black}
\begin{abstract}
	The quest for optimal operation in environments with unknowns and uncertainties is highly desirable but critically challenging across numerous fields. This paper develops a dual control framework for exploration and exploitation (DCEE) to solve an auto-optimisation problem in such complex settings. In general, there is a fundamental conflict between tracking an unknown optimal operational condition and parameter identification. The DCEE framework stands out by eliminating the need for additional perturbation signals, a common requirement in existing adaptive control methods. Instead, it inherently incorporates an exploration mechanism, actively probing the uncertain environment to diminish belief uncertainty. An ensemble based multi-estimator approach is developed to learn the environmental parameters and in the meanwhile quantify the estimation uncertainty in real time. The control action is devised with dual effects, which not only minimises the tracking error between the current state and the believed unknown optimal operational condition but also reduces belief uncertainty by proactively exploring the environment. Formal properties of the proposed DCEE framework like convergence are established. A numerical example is used to validate the effectiveness of the proposed DCEE. Simulation results for maximum power point tracking are provided to further demonstrate the potential of this new framework in real world applications.
\end{abstract}
}

{\bf \emph{Note to Practitioners}---In numerous engineering applications, it is highly desirable to operate a system to improve the efficiency, enhance performance or save energy. However, attaining this optimal control is a challenging task, due to the presence of unknown system and/or environment parameters. We develop a principled approach to balance between exploration and exploitation, involving active learning to estimate unknown parameters and tracking the optimal operational condition based on current estimation. This paper provides a unified framework to solve general auto-optimisation control problems. The simulation results demonstrate that the proposed method outperforms existing methods in terms of efficiency and optimality for maximum power point tracking problem, and it can be readily implemented for many other engineering problems. Future research include generalising the proposed method to nonlinear systems, as well as exploring novel applications to facilitate the widespread adoption of our method.}


\section{Introduction}\label{sec: Introduction}
Traditionally, adaptive control algorithms are mostly designed for either regulation problems with known setpoints or tracking problems with known reference trajectories \cite{zheng2022quasi}. In many applications, setpoints or references are usually dependent on unknown or changing environment parameters, and thus cannot be pre-specified in advance. Operating a system at optimal condition is strongly desirable for best profit, productivity or efficiency, but it can be particularly challenging in an unknown or changing environment  due to the presence of uncertainties, disturbances and noises. Typical examples include anti-lock braking systems to maintain maximal friction under various unknown road surfaces and vehicle conditions \cite{zhang2007numerical},  maximum power point tracking to continuously deliver the highest possible power to the load in presence of variations in environments \cite{leyva2006mppt, lekube2016rotational}. In this paper, these examples are referred to as auto-optimisation control problems.

As a classic control problem with a wide range of applications, early solution for static optimal operation can be traced as far back as 1922 \cite{tan2010extremum}. 
It was popular in 1950s and 1960s, and regained significant attention since 2000s due to a solid theoretical foundation established for the stability and performance in \cite{krstic2000stability,krstic2000performance}.
Several approaches have been proposed under different names including extremum seeking control \cite{guay2003adaptive, krstic2000stability} and hill-climbing systems \cite{sullivan2001convergence}. 
In the literature, the concept of self-optimisation control is introduced as a distinct control design methodology \cite{skogestad2000plantwide}. It centers on identifying a physical or virtual variable that remains constant despite disturbances or uncertainties when the system reaches its optimal state. This concept slightly differs from auto-optimisation control, which aims to maintain the system at an optimal setpoint. This setpoint optimises a performance function that may vary with unknown or changing environmental parameters, effectively adapting to uncertainties, disturbances and noises.
Since the optimal operation is unknown and possibly changes during the operation, a control system must be able to adapt to unknown or changing environments, for example, by means of learning, adaptation and action through limited interactions between the system and its operational environment. Then, the control system devises possible strategies to track the estimated setpoints or references based on its perceived environment knowledge and the level of confidence.
This type of extremum seeking and learning problem has been identified one of the key emerging methodologies in ``Control for societal-scale challenges: Road map 2030" in the field of control systems \cite{alleyne2023control}. Additionally, an optimal balance between exploration and exploitation is deemed as a crucial aspect for learning and control problems, which we will discussed in the sequel.  

Generally speaking, there are dual objectives in an auto-optimisation control problem in an unknown and uncertain environment: \emph{parameter identification} and \emph{optimality tracking}. Quite often, the dual objectives are conflicting in the sense that new observations do not provide sufficient information for identifying the unknown parameters when the system state settles to some local optimal solutions. This phenomenon widely exists in adaptive extremum seeking when an extreme searching algorithm converges to its local optimal solution, the identifiability will naturally loss due to the lack of persistent excitation (PE). As a trade-off, dither perturbations are introduced on purpose to sustain the identifiability, but such dithers inevitably deteriorate the tracking performance. Various approaches have been proposed to design the dither signals, e.g., sinusoidal perturbations \cite{krstic2000stability,tan2006non}, stochastic perturbations~\cite{manzie2009extremum,liu2010stochastic} and decaying perturbations~\cite{xie2022adaptive}. 
However, they are usually pre-specified, and thereby cannot make online adjustments according to real-time inference performance. In other words, active learning cannot be embedded, that is, actively generate data for the purpose of learning.

This paper proposes a new approach to auto-optimisation control by embedding \emph{active learning} from a new perspective: dual control of exploration and exploitation (DCEE). DCEE was originally proposed in \cite{Chen2021DCEE} for autonomous search of sources of atmospheric release where the source location and other environmental factors are unknown.  To realise autonomous search, it proposes each move of the robotic agent shall have dual effects: driving the agent towards the believed location of the source (exploitation) and probing the environment to reduce the level of uncertainty of the current belief (exploration). An optimal autonomous search strategy is realised by optimally trading-off these two effects. We argue in this paper that DCEE is actually applicable to a much wider range of systems that operate in an unknown or uncertain environment without well-defined control specifications, e.g., the reward or cost functions are unknown. We present a new auto-optimisation control framework by extending DCEE from a specific autonomous search application to a general design approach for achieving or maintaining optimal operation in an unknown environment. 
Furthermore, the DCEE scheme developed in this paper is different from the classic dual control in handling the two intricate coupling elements of the \emph{system} and the \emph{environment}. Existing dual control approaches impose a probing effect on the \emph{system} itself, for example, state estimation in stochastic control \cite{mesbah2018stochastic, chen2018approximating} and parameter estimation in adaptive control \cite{bugeja2009dual, filatov2000survey}. On the other hand, the dual effect introduced in our formulation is used to explore the operational \emph{environment}, as our objective is to acquire a better understanding of the unknown environment such that the agent is able to approach the true optimal operational condition.  This exploration approach bears resemblance to the mechanisms encountered in reinforcement learning, where the focus is also on the interactive interplay between the system and its environment \cite{Li2023AID-RL,chai2022deep}.

The contribution of this paper is of twofold. On one side, for auto-optimisation control problems, we propose a new and systematic framework which is able to actively probe the environment to reduce the level of uncertainty through active learning.
 There is no need to artificially introduce perturbation as in the current extremum seeking control. It also provides an optimal transition from any initial operation condition to acquire the unknown optimal operation condition in terms of a reformulated objective conditional upon current knowledge and future predicted information. By formulating the auto-optimisation control in this framework, it enables to establish proven properties by getting access to a wide range of theoretic tools in control theory such as parameter adaptation and optimal control. On the other side, we generalise and extend the DCEE concept from a specific application, where specific system dynamics, reward function and properties are considered, to a general control system problem. A systematic design procedure for general descriptions of the system and control objectives is presented. We show that DCEE provides a powerful and promising framework to design control systems operating in an uncertain environment, which is an important feature of autonomous systems.  

Compared with all the existing schemes for auto-optimisation control, our approach is most related to the work where the model based approach is adopted and the uncertainty of the objective or system dynamics are parameterised by uncertain parameters \cite{guay2003adaptive, gauy2017proportinal, krstic2000stability, skogestad2000plantwide, mesbah2018stochastic}. There are three main features in the new DCEE based auto-optimisation control framework, detailed as follows. 

\begin{enumerate}
	\item The proposed method embeds an active learning effect allowing the system to actively explore the unknown environment to reduce the level of uncertainty. Instead of using computationally expensive particle filters in information-driven methods, this paper develops an efficient multi-estimator based ensemble approach to quantify the estimation uncertainty online, based on which the controller effectively trades off between \emph{exploration and exploitation} to balance the dual objectives of identification and tracking. 
	\item The ensemble based estimation method advocated in this paper is distinct from those probabilistic or dynamic ensemble estimation approaches dedicated for machine learning problems.  Existing active learning based algorithms \cite{Houthooft2016VIME,chua2018deep,Cui20221492} utilise neural networks or ensembles of randomly generated dynamic models  to acquire information about the environment, which makes it challenging to extract physically meaningful parameters for the auto-optimisation control problem. The proposed multi-estimator based ensemble approach makes use of the environment model and the learned parameters are physically meaningful in control problems. 
	\item Different from all the existing schemes where probing effect is artificially introduced or inserted (usually by means of dithers and perturbations), the probing effect \emph{naturally} occurs depending on the confidence of the estimation by assembling the outcomes of these individual estimators.
\end{enumerate}

{\color{black}
In order to guide the reader through the content, we provide a brief summary outlining the structure of this paper. In Section~\ref{sec: Problem Formulation}, we formulate the auto-optimisation control problem and demonstrate the dual effects embedded in the new formulation. In Section~\ref{sec: DCEE for Integrator}, an active learning based ensemble approach is developed for unknown environment acquisition and then a dual controller for exploration and exploitation is designed to achieve optimal trade-off between parameter identification and optimality tracking for a special single integrator system. Section~\ref{sec: DCEE for Linear Systems} extends DCEE to general linear systems and formal properties of the proposed auto-optimisation control method are established. Section~\ref{sec: Numerical Example} demonstrates the effectiveness of the proposed algorithm using a numerical example. Section~\ref{sec: MPPT} formulates maximum power point tracking (MPPT) problem as an auto-optimisation control problem and compares the proposed algorithm with other existing approaches. Section~\ref{sec: Conclusion} concludes this paper. 
}

\section{Problem Statement}\label{sec: Problem Formulation}
In this section, we elaborate the dual effects embedded in the reformulated auto-optimisation control problem. Then, an ensemble active learning based approach is introduced to realise efficient parameter adaptation and assess the estimation performance. 

\subsection{Dual Control Reformulation}
Consider a reward function for a system operating in an unknown environment
\begin{equation}\label{eqn: reward function}
	J(\theta,y) = \phi^{\tp}(y) \theta
\end{equation}
where $\theta  \subset \bR^m $ is unknown, depending on the operational environment, $y\in \bR^q$ is the system output, and $\phi(y) \in \bR^m$ is the basis function of the reward function. In other words, the reward function is parameterised by unknown $\theta^*$. 
{\color{black}
The reward function \eqref{eqn: reward function} accommodates a broad spectrum of functions through either first-principle modelling or function approximation methods. Specifically, first-principle modeling can be used to construct various engineering objectives, such as quadratic or polynomial functions with unknown coefficients. Moreover, it can also encompass neural network approximations. In these cases, the neural network's activation functions, like radial basis function (RBF) networks or Gaussian Kernels, act as the basis $\phi(y)$, and the network's weights correspond to the unknown parameters $\theta$. Therefore, the considered reward function demonstrates wide applicability in a range of scenarios.}

Without loss of generality, it is assumed the the optimal condition is achieved at the maximum of $J$. 
An auto-optimisation control is designed to automatically drive the system to the unknown operational condition, maintain there despite disturbances and automatically adjust the optimal operation condition accordingly when the operational environment changes. 

The system dynamics under consideration are described by
\begin{equation}\label{system dyanmics}\begin{aligned}
		& x(k+1)=Ax(k)	+ Bu(k) \\
		& y(k) = Cx(k)
	\end{aligned}
\end{equation}
where $x(k)\in \BR^n$, $u(k) \in \BR^p $ and $y(k) \in \BR^q$ are system state, control input and output, respectively, and $ A \in \bR^{n\times n}, B\in \bR^{n\times p}, C \in \BR^{q\times n} $ are constant matrices. 
Suppose that at each time, the system output and the reward $J(k)$ can be measured or derived subject to measurement noise $v(k)$. The measurement information at the $k$th step is given by  
\begin{equation}\label{eqn: information with noise}
	z(k)=[x(k); y(k);J(k)+v(k)]
\end{equation}
and the information state is denoted as
\begin{equation}
	I_k=[u(k-1);z(k)]
\end{equation}
All the measurement up to the current time $k$ is given by
\begin{equation}
	\mathbf{I}_k=[I_0,I_1, \ldots,I_k]
\end{equation}
with $I_0=[z(0)]$.

There are two ways to formulate this problem using the dual control for exploration and exploitation (DCEE) concept. The first approach is similar to extremum seeking control \cite{guay2003adaptive, gauy2017proportinal} aiming to select the control such that the reward function is maximised with all the information up to now including the prior and all the measurements
\begin{equation}\label{eqn: dual formulation 1}
	\max_{u(k) \in \BR^p} \bE_{\theta,I_{k+1|k}}\{J(\theta,y(k+1|k))|\mathbf{I}_{k+1|k}\}
\end{equation}
subject to the system dynamics \eqref{system dyanmics}, where $\mathbf{I}_{k+1|k}=[\mathbf{I}_k, I_{k+1|k}]$ with $I_{k+1|k} = [u(k),z(k+1|k)]$. $z(k+1|k)$ consists of the predicted output $y(k+1|k)$ and the predicted reward function under the control $u(k)$. 

Another approach is to drive the system output to the unknown optimal condition directly. Since the optimal operation condition is unknown, the best one can do is to drive the system to the best estimation of the optimal operation condition with all the information up to now. This can be formulated as
\begin{equation}\label{eqn: optimal operational condition formulation}
	\min_{u(k) \in \BR^p} \bE\{(y(k+1|k)-r^*)^{\tp}(y(k+1|k)-r^*)|\mathbf{I}_{k+1|k}\}
\end{equation}
where $ r^* = l(\theta^*) $ denotes the predicted optimal operational condition conditional upon $\mathbf{I}_{k+1|k}$, which is a function of  the environment parameter $\theta^*$. In realm of auto-optimisation control, it is often required that the mapping $l(\theta)$ is a smooth function of $\theta$ and $r^* = l(\theta^*)$ is a unique optimum of the objective function~\cite{krstic2000stability}. 

These two problems have been solved previously in autonomous search \cite{Chen2021DCEE,Li2021CLEE}. The research question is how to extend these results from this specific application to general auto-optimisation control problems. In this paper, we will focus our attention on the latter formulation in \eqref{eqn: optimal operational condition formulation}, which is related to the operational condition determined by unknown environment parameters. 

Before proceeding further, we demonstrate that the control input $u(k)$ obtained by minimising \eqref{eqn: optimal operational condition formulation} \emph{naturally} carries dual effects corresponding to exploration and exploitation, respectively. Intuitively, the control input $u(k)$ will influence the future system state $ y(k+1|k) $ via the system dynamics \eqref{system dyanmics}, and at the same time affect the future information to be collected $I_{k+1|k}$ via the reward function in \eqref{eqn: reward function} from the environment subject to uncertainties.

We define the predicted nominal operational condition as 
\begin{equation}\label{eqn: nominal J}
	\bar r(k+1|k)  = \bE \left[ r(k+1|k)\vert \mathbf I_{k+1 | k} \right]
\end{equation}
based on which the prediction error conditional on $ \mathbf I_{k+1 | k}$ can be written as 
\begin{equation}\label{eqn: residual J}
	\tilde r(k+1|k) = r^*- \bar r(k+1|k) . 
\end{equation}
Expanding \eqref{eqn: optimal operational condition formulation} and substituting \eqref{eqn: nominal J} and \eqref{eqn: residual J} into \eqref{eqn: optimal operational condition formulation}, we have 
\begin{equation}\begin{aligned}\label{eqn: dual derivation}
		\bE & \left[    \| y(k+1|k) - \bar r(k+1|k)  -  \tilde r(k+1|k) \|^{2} \vert \mathbf I_{k+1 | k} \right] \\
		= & \bE \left[    \| y(k+1|k) - \bar r(k+1|k)  \|^{2} \vert \mathbf I_{k+1 | k} \right] \\
		& - 2\bE \left[    ( y(k+1|k) - \bar r(k+1|k))^{\tp}  \tilde r(k+1|k)  \vert \mathbf I_{k+1 | k}  \right]  \\ 
		&  + \bE \left[  \|\tilde r(k+1|k) \|^2  \vert \mathbf I_{k+1 | k}     \right].
	\end{aligned}
\end{equation}
It follows from the definition of $ \tilde r(k+1|k)$ in \eqref{eqn: residual J} that $\bE \left[  \tilde r(k+1|k)  \vert \mathbf I_{k+1 | k}  \right] =0$. Thus, by further noting that  $y(k+1|k)$  and $ \bar r(k+1|k) $ are deterministic, the cross term in \eqref{eqn: dual derivation} equals to zero, yielding
\begin{equation}\begin{aligned}\label{eqn: dual derivation 2}
		D(u(k)):= & \bE \left[    \| y(k+1|k) - \bar r(k+1|k)  \|^{2} \vert \mathbf I_{k+1 | k} \right] \\
		&  + \bE \left[   \| \tilde r(k+1|k) \|^2   \vert \mathbf I_{k+1 | k}     \right].
	\end{aligned}
\end{equation}

\begin{remark}
	The objective function in \eqref{eqn: dual derivation 2} exhibits dual effects. Minimising the first term in \eqref{eqn: dual derivation 2} drives the system state to estimated nominal value, which corresponds to the exploitation effect. In control terminology, it can be understood as tracking a nominal reference, thus also referred to as optimality tracking. The second term characterises the level of uncertainty (variance in this case) associated with the predicted optimal operational condition, which is related to the exploration effect. According to the classic dual control concept \cite{feldbaum1960dual1,bar1974dual}, a control input is said to have dual effects if it can affect at least one $j$th-order central moment of a state variable ($j>1$), in addition to its effect on the state.   
 {\color{black}
 In fact, the dual control framework developed in this paper is a generalisation of the classic one \cite{feldbaum1960dual1} in the sense that our formulation deals with not only system uncertainty but also environment uncertainty (the operational condition $r^* = l(\theta^*)$ is determined by the environment parameters $\theta^*$). This subtle difference endows the system with capability of exploring the operational environment and in the meanwhile exploiting its current belief. Moreover, this paper provides a solution framework to auto-optimisation control problems with complete theoretical analysis, whereas \cite{feldbaum1960dual1} considers adaptive control problems without formal analysis.}
 Recently, DCEE has demonstrated superior and promising performance in autonomous search \cite{Chen2021DCEE,Rhodes2021dual}.
\end{remark}

\begin{remark}
	According to \cite{Antsaklis2020autonomy}, the level of autonomy can be measured in terms of the set of goals that the system is able to accomplish subject to a set of uncertainties.
	As a result, it is required that the system can exploit its available knowledge to accomplish the goals, and at the same time it should be able to actively explore the operational environment to reduce knowledge uncertainty. Effective trading-off between exploration and exploitation has been a long standing issue, particularly in artificial intelligence, control and decision-making in complex and uncertain environment.  In control society, some recent works explicitly introduce trade-off coefficients to incorporate the exploration terms into model predictive control problems, e.g., \cite{Heirung2017dual, mesbah2018stochastic}. This inevitably incurs tedious efforts in tuning the coefficients to balance exploration and exploitation. In view of the derivation of \eqref{eqn: dual derivation 2}, it is clear that the dual effects in DCEE are naturally embedded, since they are derived from a physically meaningful value function in~\eqref{eqn: optimal operational condition formulation}.
\end{remark}

\subsection{Ensemble based Active Learning}
Efficient gradient descent algorithms can be used to estimate the unknown parameters. The performance of single estimator based optimisation algorithm is quite poor, due to noisy measurement and nonlinear modelling (see examples in autonomous search \cite{Hutchinson2017review,Li2021CLEE}). Recently, the ensemble-based approximation in machine learning community has demonstrated great success with tractable computational load \cite{chua2018deep, Lakshminarayanan2017simple}. In this paper, we develop a multi-estimator based learning method for parameter adaptation, which shows comparable performance as particle filter using much less computational resources in autonomous search application \cite{Li2021CLEE}.

Considering an ensemble of $N$ estimators, the dual formulation in \eqref{eqn: dual derivation 2} becomes 
\begin{equation}\begin{aligned}\label{eqn: dual derivation 3}
		\min_{u(k) \in  \BR^p} 	&\  D(u)  = \| y(k+1|k) - \bar r(k+1|k)  \|^{2}  + \mathcal P(k+1|k)\\
		\text{subject to} &\ x(k+1|k)=Ax(k)+Bu(k) \\
		& \  y(k+1|k) = Cx(k+1|k)
	\end{aligned}
\end{equation}
where the nominal estimate and variance of the estimated optimal condition are drawn from the ensemble, i.e., 
\begin{align}
	\bar{r}(k+1 | k) =&  \frac 1N \sum_{i=1}^N{{r}}_i(k+1 | k)  =  \frac 1N \sum_{i=1}^N{l}(\theta_i(k+1 | k)) \\
	\mathcal P(k+1| k) =& \frac 1N \sum_{i=1}^N  ( r_i(k+1|k)- \bar{ r}(k+1|k) )^{\tp} \nonumber  \\  & \times ( r_i(k+1|k)- \bar{ r}(k+1|k) )  
\end{align}
where the subscript $i\in \mathcal N$ denotes the index of the estimators, with $\mathcal N$ representing the set of the ensemble.
Note that the relationship between the predicted optimal condition and the unknown parameter, i.e., $r_i(k+1 | k)  =  l(\theta_i(k+1 | k))$, is usually known. For example, in autonomous search application, $\theta^*$ is composed of the unknown source location and other environment parameters, like wind direction and wind speed.
%, and the optimal operation condition $r^*$, i.e., source location, is simply part of $\theta^*$. 
The optimal operation condition $r^*$ in autonomous search is the source location, i.e., part of $\theta^*$, which serves as a tracking reference for the search agent.

In order to estimate the unknown parameter $\theta^*$, we apply a gradient-descent regression method  \cite{ortega2020modified}, designed as
\begin{equation}\label{eqn: environment learning}
	\begin{aligned}
		\theta_{i}(k) =  & \theta_{i}(k-1)   -   \eta_i\phi(y(k-1)) \\ & \times \left[  \phi(y(k-1))^{\tp}  \theta_i(k-1) -J(k-1) \right] , \ \forall i \in \mathcal N.
	\end{aligned}
\end{equation}
where $\eta_i >0$ is the learning rate of the $i$th estimator; $J(k-1)$ denotes the observed reward with measurement noise in (\ref{eqn: information with noise}) at $y(k-1)$; and $ \theta(k)$ denotes the estimate of unknown reward parameter $\theta^*$. The estimators are randomly initialised 
or they can be initialised according to \emph{a priori} pdfs of the unknown parameters if available.
Denote the estimation error as $\tilde \theta_{i}(k) = \theta_i (k) - \theta^* $. Then, by noting $J(k-1) = \phi(y(k-1))^{\tp} \theta^* + v(k-1) $, we have 
\begin{equation}\label{eqn: mutli-estimator error dynamics}
	\begin{aligned}
		\tilde \theta_{i}(k) =  &  \left[ I_{m}  -   \eta_i\phi(y(k-1))  \phi(y(k-1))^{\tp} \right] \tilde \theta_i(k-1) \\
		&  -   \eta_i \phi(y(k-1)) v(k),
		\ \forall i \in \mathcal N.
	\end{aligned}
\end{equation}
Denoting the extended parameter error as $\tilde \Theta(k) = \col\{\tilde \theta_1(k), \dots, \tilde \theta_N(k) \}$, where $ \col \{\cdot\}$ denotes a column vector formed by stacking the elements on top of each other,  \eqref{eqn: mutli-estimator error dynamics} can be written in a compact form as
\begin{equation}\label{eqn: mutli-estimator error dynamics compact form}\begin{aligned}
		\tilde \Theta(k) = & \big[ I_N \otimes  \left( I_{m}  -   \eta_i\phi(y(k-1))  \phi(y(k-1))^{\tp} \right) \big ]\tilde \Theta(k-1) \\
		& -  \big[ I_N \otimes \eta_i \phi(y(k-1)) \big ] (1_N \otimes v(k-1) ).
	\end{aligned}
\end{equation}

In an ensemble-based adaptation, we take their average as the current estimation of the unknown parameters. Thus, averaging \eqref{eqn: mutli-estimator error dynamics compact form}, we have 
\begin{equation}\label{eqn: mutli-estimator sum error dynamics} 
	\begin{aligned}
		\tilde{\Theta}_{av} (k) = &   \frac 1N  (1_N^{\tp}  \otimes I_m) {\tilde \Theta}(k) \\ = &  \frac 1N  (1_N^{\tp}  \otimes I_m)   \left[I_N \otimes ( I_{m}  -   \eta \phi \phi^{\tp} ) \right] \tilde \Theta(k-1)  \\ 
		& - \frac 1N  (1_N^{\tp}  \otimes I_m) \big[ I_N \otimes \eta \phi \big ] (1_N \otimes v(k-1) )
		\\ = &    \frac 1N  (1_N^{\tp}  \otimes I_m)  \tilde \Theta(k-1)  - \frac 1N  (1_N^{\tp}  \otimes \eta \phi \phi^{\tp} ) \tilde \Theta(k-1) \\
		& -\eta \phi v(k-1).
	\end{aligned}
\end{equation}

\begin{remark}
An important observation is that even though we have the same regressor $\phi$ at one time instant, its excitation impact to each estimator will be different since $\phi \phi^{\tp} \tilde{\theta}_i \neq \phi \phi^{\tp} \tilde{\theta}_j $, $\forall i\neq j $, almost surely. 
Due to the introduction of parameter extension by multiple estimators, at any time instant the average estimation can always be excited when there are sufficient estimators. In addition, by introducing a group of estimators, it is possible to evaluate and make full use of the estimation uncertainty by sampling the outcomes of the ensemble in an online manner, which is proved to be crucial in DCEE \cite{Chen2021DCEE} as we will discuss in the sequel. Another desirable feature of the ensemble approach is its resilience to measurement noises. In view of the last term in \eqref{eqn: mutli-estimator sum error dynamics}, instantaneous noises will be averaged out under multiple estimators such that the overall performance of the ensemble can be improved. 
\end{remark}

\section{DCEE for Single Integrator}\label{sec: DCEE for Integrator}

\subsection{Algorithm Development}\label{subsec: Algorithm Development}
In high-level decision-making, system behaviours are usually simplified as single integrators by ignoring low-level dynamics. In this paper, we begin with DCEE for this special case
\begin{equation}\label{eqn: single integrator}
y(k+1) = y(k) + u(k).
\end{equation} 
%by letting $CA=I$ and $CB=I$. 
For general linear systems, we will use this as an internal reference generator, as will be shown later in Section \ref{sec: DCEE for Linear Systems}.

With the estimated environment parameter in \eqref{eqn: environment learning}, the dual controller can be designed as 
\begin{equation}\label{eqn: dual controller}\begin{aligned}
	& y(k+1)=y(k)	 + u(k) \\
	& u(k) = - \delta_k \big[   \nabla_{y} \mathcal C(k+1|k) +  \nabla_{y} \mathcal P(k+1|k)   \big]
\end{aligned}
\end{equation}
where $\mathcal C(k+1|k)= \| y(k) -\bar{r}(k+1 | k)  \| ^2 $ denotes the exploitation term, and $ \mathcal P(k+1|k)$ is the exploration term  in the dual objective \eqref{eqn: dual derivation 3}. 
According to the gradient-descent regression in \eqref{eqn: environment learning}, the predicted mean of the $N$ ensemble $\theta_i(k+1|k)$, denoted as $\bar \theta(k+1|k)$, is given by 
\begin{equation}\begin{aligned}
	\bar \theta(k+1|k) = & \frac 1N \sum_{i=1}^{N} \theta_i(k+1|k) \\
	= & \frac 1N \sum_{i=1}^{N} (\theta_i(k) - \eta_i F_i (k+1|k) )
\end{aligned}
\end{equation}
where 
\begin{equation}\begin{aligned}
	F_i (k+1|k) 
	= & [J( \theta_i (k), y(k) )-J(k+1|k) ]\phi (y(k)) 
\end{aligned}
\end{equation}
with $J(k+1|k)$ being the predicted future reward based on current belief $\{ \theta_i(k), \forall i\in \mathcal N\} $. Note that the predicted future reward is noise-free as there is no influence from sensory devices in prediction. In this paper, we use the average of $\theta_i(k), \forall i\in \mathcal N$ to evaluate the predicted future reward, i.e., $J(k+1|k) = J(\bar \theta(k), y(k))$. Similarly, the predicted variance of the ensemble is given by 
\begin{equation}\begin{aligned}
	\mathcal P(k+1|k) =\operatorname{trace}(  \bm F(k+1|k)^{\tp}{ \mathcal P}(k|k) \bm F(k+1|k))
\end{aligned}
\end{equation}
where 
\begin{equation}\begin{aligned}
	\bm F(k+1|k) = & \col \{F_1 (k+1|k), \dots, F_N (k+1|k) \} \\
	\mathcal P(k|k) = & \operatorname{cov} \{ \theta_i(k), \forall i\in \mathcal N\}  \\ = & \diag \{ ( \theta_{1}(k) - \bar{\theta}(k) )( \theta_{1}(k) - \bar{\theta}(k) )^{\tp}, \dots, \\ & ( \theta_{N}(k) - \bar{\theta}(k) ) ( \theta_{N}(k) - \bar{\theta}(k) )^{\tp}  \}
\end{aligned}
\end{equation}
where $\operatorname{cov}\{\cdot\}$ is a covariance operator evaluating the covariance matrix of the ensemble, and $\diag \{\cdot\}$ denotes a  block-diagonal matrix by putting the elements on its main diagonal. 
Using the predicted mean $\bar \theta(k+1|k) $ and the predicted covariance $\mathcal P(k+1|k)$ of the unknown environmental parameter, the dual control terms in \eqref{eqn: dual controller} can be obtained.

\subsection{Convergence Analysis}\label{subsec: Convergence Analysis}
In this section, we will examine the convergence of the proposed dual control algorithm, by leveraging parameter adaptation and optimisation techniques. To this end, we introduce some fundamental assumptions that will be used to facilitate the convergence analysis of the proposed dual control algorithm.

\begin{assumption}\label{asm: PE}
There exist positive constants $T \in \mathbb Z^+$ and $\gamma \geq \beta >0$ such that 
\begin{equation}\label{eqn: PE condition}
	\gamma I_m \geq \sum_{k=t}^{t+T} [\phi(y(k))]  [\phi(y(k))]^{\tp}  \geq \beta I_m >0,\ \forall t>0.
\end{equation}
\end{assumption}

\begin{assumption} \label{asm: noise condition}
The measurement noise $v(k)$ is independent and identically distributed with bounded variance, i.e., 
\begin{equation}\label{eqn: noise v}
	\begin{aligned}
		\mathbb{E}\left[v(k) \right] &=0     \\
		\mathbb{E}\left[ \|v(k)\|^2 \right] & \leq  
		{\varrho}^{2}.
	\end{aligned}
\end{equation}
\end{assumption}

\begin{assumption}\label{asm: reward convexity}
The reward function $J(\theta,y)$ is twice differentiable and strictly concave on $y$ for any $\theta \in \bR^m$, that is,  
\begin{equation}\label{eqn: strong convexity}
	\frac{\partial^2 J(\theta,y)}{\partial y^2 }  > 0. 
\end{equation}
\end{assumption}

\begin{remark}\label{rem: 3}
Assumption~\ref{asm: PE} is a standard persistent excitation (PE) condition to ensure the identifiability of the unknown environmental parameter $\theta$. Extensive techniques on parameter adaptation have been reported in the past few decades aiming at relaxing or fulfilling the conditions of PE \cite{ortega2020modified,guay2003adaptive}. If we introduce a memory-based regressor extension to the parameter adaptation algorithm in \eqref{eqn: environment learning}, the PE condition can be relaxed to interval excitation \cite{ortega2020modified}.  Assumption~\ref{asm: noise condition} implies that the noises imposed on sensory information are unbiased with bounded variances \cite{Li2021CLEE}. Assumption~\ref{asm: reward convexity} guarantees the existence and uniqueness of the optimal operational condition, i.e., $ r^*=l(\theta^*)$, which is widely used in adaptive control and extremum seeking control \cite{guay2003adaptive, adetola2007parameter}. Note that the mapping between the optimal operational condition and parameter $\theta$ can be obtained by solving $\frac {\partial J(\theta, y )} {\partial y} = 0 $.
\end{remark}

First, we examine the convergence of the gradient-descent regression method in \eqref{eqn: environment learning}.
\begin{theorem}\label{thm: 1}
Under Assumptions \ref{asm: PE} and \ref{asm: noise condition}, there exists a constant $\eta^* >0$ such that, for any $0<\eta_i< \eta^*$, the estimates, $\hat \theta_i(k),\ \forall i\in \mathcal N, $ converge to a bounded neighbourhood of the true environmental parameter $\theta^*$. Moreover, the mean-square-error of the estimator is convergent and bounded by 
\begin{equation}
	\bE \| \tilde \theta_{i}(k) \|^2  \leq \frac{\eta_i^2 L^2 \varrho^2}{1- \max_{j\in\{1,\dots, k-1\}} \|A_i(j)\|} 
\end{equation}
where $A_i (j) =  [I_{m}  -   \eta_i[\phi(y(j))]  [\phi(y(j))]^{\tp}]^{\tp} [I_{m}  -   \eta_i[\phi(y(j))]  [\phi(y(j))]^{\tp}] $ and $L$ denotes the bound of the regressor $\phi$. Moreover, in absence of measurement noises, $ \lim_{k\rightarrow \infty} \bE \| \tilde \theta_{i}(k) \|^2 = 0$.
\end{theorem}
\begin{proof}
In view of \eqref{eqn: mutli-estimator error dynamics} and Assumption~\ref{asm: noise condition}, the expectation of the estimate is given by 
\begin{equation}\label{eqn: expectation of estimation error}
	\begin{aligned}
		\bE[ \tilde \theta_{i}(k) ] =  &  \left[ I_{m}  -   \eta_i[\phi(y(k-1))]  [\phi(y(k-1))]^{\tp} \right] \bE[  \tilde \theta_i(k-1)] \\ &
		\ \forall i \in \mathcal N.
	\end{aligned}
\end{equation}
According to Assumption~\ref{asm: PE}, there exists a constant $\eta^*$ such that, for any $0<\eta_i<\eta^*$, $0<\eta_i[\phi(y(k-1))]  [\phi(y(k-1))]^{\tp}< I_m$. Consequently, for any $0<\eta_i<\eta^*$, we have 
\begin{equation}
	0< I_{m}  -   \eta_i[\phi(y(k-1))]  [\phi(y(k-1))]^{\tp} <I_m.
\end{equation}
It follows from \eqref{eqn: expectation of estimation error} that 
\begin{equation}\label{eqn: norm of expected estimation error}
	\begin{aligned}
		\| \bE[ \tilde \theta_{i}(k) ] \| \leq   &  \left\| I_{m}  -   \eta_i[\phi(y(k-1))]  [\phi(y(k-1))]^{\tp} \right\| \\ & \times \| \bE[  \tilde \theta_i(k-1)] \|,
		\ \forall i \in \mathcal N.
	\end{aligned}
\end{equation}
Therefore, 
\begin{equation}\label{eqn: norm of expected estimation error 2}
	\begin{aligned}
		\| \bE[ \tilde \theta_{i}(k) ] \| \leq   & \prod_{j=1}^{k}  \left\| I_{m}  -   \eta_i[\phi(y(j-1))]  [\phi(y(j-1))]^{\tp} \right\| \\ & \times \| \bE[  \tilde \theta_i(0)] \|,
		\ \forall i \in \mathcal N.
	\end{aligned}
\end{equation}
For any bounded error $\tilde \theta_i(0)$, the expectation of the estimator converge to zero. 

Moreover, the variance of the estimators can be bounded under Assumption~\ref{asm: noise condition}. Taking the squared Euclidean norm of \eqref{eqn: mutli-estimator error dynamics} yields 
\begin{equation}\label{eqn: squared Euclidean norm}
	\begin{aligned}
		\| \tilde \theta_{i}(k) \|^2 =  &  \big\| \big[ I_{m}  -   \eta_i[\phi(y(k-1))]  [\phi(y(k-1))]^{\tp} \big] \tilde \theta_i(k-1) \big \| ^2 \\
		&  + \|   \eta_i\phi(y(k-1))v(k)\|^2 \\
		& -   2 \big[ [ I_{m}  -   \eta_i[\phi(y(k-1))]  [\phi(y(k-1))]^{\tp} ]\\ & \times  \tilde \theta_i(k-1)\big]  \left[ \eta_i\phi(y(k-1)) v(k) \right],
		\ \forall i \in \mathcal N.
	\end{aligned}
\end{equation}
Applying expectation operation to \eqref{eqn: squared Euclidean norm} leads to 
\begin{equation}\label{eqn: squared Euclidean norm 2}
	\begin{aligned}
		\bE \| \tilde \theta_{i}(k) \|^2 =  & \bE \big\| \left[ I_{m}  -   \eta_i[\phi(y(k-1))]  [\phi(y(k-1))]^{\tp} \right]  \\ &  \times \tilde \theta_i(k-1) \big \| ^2 \\
		&  + \bE \|   \eta_i\phi(y(k-1))v(k)\|^2 ,
		%		& -   2 \big[ [ I_{m}  -   \eta_i[\phi(y(k-1))]  [\phi(y(k-1))]^{\tp} ] \tilde \theta_i(k-1)\big] \\ & \times  \left[ \eta_i\phi(y(k-1)) v(k) \right],
		\ \forall i \in \mathcal N.
	\end{aligned}
\end{equation}
where $\mathbb{E}\left[v(k) \right] =0 $ has been used to eliminate the cross term.   
Denoting $A_i (k-1) =  [ I_{m}  -   \eta_i[\phi(y(k-1))]  [\phi(y(k-1))]^{\tp}]^{\tp} [ I_{m}  -   \eta_i[\phi(y(k-1))]  [\phi(y(k-1))]^{\tp} ]$ and applying the variance bound in \eqref{eqn: noise v}, we have 
\begin{equation}\label{eqn: expected bound of estimation error}
	\bE \| \tilde \theta_{i}(k) \|^2  \leq  \bE \big\| \tilde \theta_i(k-1) \big \|_{A_{i}(k-1)} ^2 + \eta_i^2 L ^2 \varrho^2.
\end{equation}
%where $L$ denotes the bound of the regressor $\phi$. 
For any $0<\eta_i<\eta^*$, the mean-square-error of the estimator is convergent and bounded by 
\begin{equation}
	\lim_{k\rightarrow \infty}\bE \| \tilde \theta_{i}(k) \|^2  \leq \frac{\eta_i^2 L^2 \varrho^2}{1- \max_{j\in\{1,\dots, k-1\}} \|A_i(j)\|} .
\end{equation}
In absence of measurement noise $v(k) = 0$, $ \lim_{k\rightarrow \infty} $ $\bE \| \tilde \theta_{i}(k) \|^2 = 0$.
This completes the proof. 
\end{proof}

\begin{remark}
Theorem~\ref{thm: 1} establishes the convergence of the estimators. The parameter adaptation algorithm together with its convergence analysis under measurement noises forms a new feature of this paper since existing studies mainly focus on noise-free scenarios \cite{ding2007performance, ortega2020modified}.   As having been discussed in Remark~\ref{rem: 3}, PE is a standard and commonly-used condition to guarantee the convergence of parameter estimators. Despite significant research efforts have been dedicated to explore weak/alternative assumptions, very few result has been obtained (see recent survey in \cite{ortega2020modified}). In the proposed dual controller \eqref{eqn: dual controller}, a probing effort is inherently embedded aiming to reduce the estimation uncertainty. Such an exploration effect from active learning is beneficial to environment acquisition, which has been validated in autonomous search application \cite{Chen2021DCEE, Li2021CLEE}. 
\end{remark}

\begin{remark}
The proposed multi-estimator assisted ensemble method for environment adaptation is a hybrid approach that combines both model-based and model-free techniques. The model-based estimators are trained according to the model structures of the reward function in \eqref{eqn: reward function}. A model-free ensemble approximation is used to estimate the mean and variance of the unknown environmental parameters in an online manner. It is widely perceived in machine learning community that model-based approach benefits from high learning efficiency due to the utilisation of model knowledge but inevitably inherits model biased errors; on the other hand, model-free approach provides a reliable way to quantify the level of estimation uncertainty but may incur additional computational burden. Recently, the hybrid method has demonstrated superior performance in simulation and experiment in machine learning due to its combined strength from both model-based and model-free learning \cite{chua2018deep, Lakshminarayanan2017simple,yue2020active}. 
Theoretical guarantee on convergence and performance of the hybrid approach has not been well-established but mainly verified by extensive simulation and experimental results. Inspired by its recent success, we develop a concurrent active learning based ensemble algorithm and establish its formal properties in this paper. Additionally, different from existing studies in active control \cite{Houthooft2016VIME, chua2018deep} where Bayesian neural networks or ensembles of dynamic models are employed to formulate information gain, the proposed DCEE method captures belief uncertainty using multi-estimator ensemble, from which physically meaningful parameters can be extracted.
\end{remark}

Denote the tracking error between current state and unknown optimal condition $r^*$ as $\tilde y(k) = y(k) - r^*$. Then, it follows from \eqref{eqn: dual controller} that 
\begin{equation}\label{eqn: error dynamics 1}\begin{aligned}
	& \tilde y(k+1)= \tilde y(k) - \delta_k \big[   \nabla_{y} \mathcal C(k+1|k) +  \nabla_{y} \mathcal P(k+1|k)   \big].
\end{aligned}
\end{equation}
Now, we analyse the convergence to the optimal operational condition. 

\begin{theorem}\label{thm: 2}
Under Assumptions \ref{asm: PE}-\ref{asm: reward convexity}, for any $0<\eta_i< \eta^*$, $y$ converges to a bounded neighbourhood of the optimal operational condition $r^*= l(\theta^*)$  if there exists a step size $\delta_k $ such that  $0< 2\| [I_n - \delta_k \mathcal L(k)] \|^2 <1$ with $  {\mathcal L}(k) =  \int_{0}^{1} \nabla_{y}^{2} \mathcal C({r^* }+\tau \tilde{y}(k) ,\bar r(k+1|k) )d \tau $.
\end{theorem}
\begin{proof}
To relate the gradient term $ \nabla_{y} \mathcal C(k+1|k) $ with $  \tilde{y}(k)$, we recall the mean value theorem~\cite{rudin1976principles}, that is, for a twice-differentiable function $ h(y): \bR^m \rightarrow \bR $, 
\begin{equation}\label{eqn: mean value theorem}
	\begin{aligned}
		\nabla h(y_1)=& \nabla h(y_2) +\left[\int_{0}^{1} \nabla^{2} h[y_2+\tau (y_1-y_2) ]d \tau \right](y_1-y_2), \\ & \forall  y_1,y_2 \in \bR^m.
	\end{aligned}
\end{equation}
Thus, we have 
\begin{equation}\begin{aligned}
		\nabla_{y}& \mathcal C (y(k),\bar r(k+1|k)) =  \nabla_{y} \mathcal C(r^*,\bar r(k+1|k))  \\ &+ \bigg[ \int_{0}^{1} \nabla_{y}^{2} \mathcal C({r^* }+\tau \tilde{y}(k),\bar r(k+1|k)  )d \tau \bigg] \tilde{y}(k)
	\end{aligned}
\end{equation}
where we have expanded the notation $\mathcal C(k+1|k)$ for clarity. Denoting $  {\mathcal L}(k) =  \int_{0}^{1} \nabla_{y}^{2} \mathcal C({r^* }+\tau \tilde{y}(k)  ,\bar r(k+1|k))d \tau $ and applying $\nabla_{y} \mathcal C(r^*,\bar r(k+1|k))=\bm 0$, we have 
\begin{equation}\label{eqn: gradient equation 1}
	\nabla_{y} \mathcal C (y(k),\bar r(k+1|k)) = \mathcal L(k) \tilde{y}(k).
\end{equation}
Applying \eqref{eqn: gradient equation 1} to \eqref{eqn: error dynamics 1} results in 
\begin{equation}\label{eqn: error dynamics 2}\begin{aligned}
		& \tilde y(k+1)= [I_n - \delta_k \mathcal L(k)] \tilde y(k) - \delta_k \nabla_{y} \mathcal P(k+1|k) . \\
	\end{aligned}
\end{equation}
To examine the boundedness of the tracking error, we take the Euclidean norm for both sides of \eqref{eqn: error dynamics 2}, yielding 
\begin{equation}\label{eqn: Euclidean norm dynamics}\begin{aligned}
		\|\tilde y(k+1)\|^2= & \| [I_n - \delta_k \mathcal L(k)] \tilde y(k) \|^2 
		%		 + \| w(k)\|^2 
		+  \| \delta_k \nabla_{y} \mathcal P(k+1|k) \|^2  \\ 
		%		 & +2 [(I_n - \delta_k \mathcal L(k))\tilde y(k)]^{\tp} w(k) \\ 
		%		 & - 2 \delta_k \nabla_{y}^{\tp} \mathcal P(k+1|k) w(k) \\ 
		& -2 \delta_k [(I_n - \delta_k \mathcal L(k))\tilde y(k)]^{\tp}  \nabla_{y} \mathcal P(k+1|k) . 
	\end{aligned}
\end{equation}
Taking the expectation of \eqref{eqn: Euclidean norm dynamics} leads to 
\begin{equation}\label{eqn: expeactation of Euclidean norm dynamics}\begin{aligned}
		\bE \|\tilde y(k+1)\|^2 \leq  & \| [I_n - \delta_k \mathcal L(k)] \|^2 \bE \| \tilde y(k) \|^2
		%		 + \bE \| w(k)\|^2 
		\\ & +  \bE \| \delta_k \nabla_{y} \mathcal P(k+1|k) \|^2  \\ 
		%		& +\bE[2 [(I_n - \delta_k \mathcal L(k))\tilde y(k)]^{\tp} w(k) ] \\ 
		%		& + \bE [ 2 \delta_k \nabla_{y}^{\tp} \mathcal P(k+1|k) w(k) ] \\ 
		&+\bE[-2 \delta_k   \nabla_{y}^{\tp}  \mathcal P(k+1|k) [(I_n - \delta_k \mathcal L(k))\tilde y(k)]]. 
	\end{aligned}
\end{equation}
The last term in \eqref{eqn: expeactation of Euclidean norm dynamics} can be written as 
\begin{equation}\label{eqn: expectation cross term}\begin{aligned}
	& \bE[ -2 \delta_k    \nabla_{y}^{\tp}  \mathcal P(k+1|k) [(I_n - \delta_k \mathcal L(k))\tilde y(k)]]  \\ 
	& \leq \| [I_n - \delta_k \mathcal L(k)] \|^2 \bE \| \tilde y(k) \|^2  +   \bE \| \delta_k \nabla_{y} \mathcal P(k+1|k) \|^2 .
\end{aligned} 
\end{equation}
Therefore, substituting \eqref{eqn: expectation cross term} into \eqref{eqn: expeactation of Euclidean norm dynamics} results in 
\begin{equation}\label{eqn: expeactation of Euclidean norm dynamics 2}\begin{aligned}
	\bE \|\tilde y(k+1)\|^2 \leq  & 2\| [I_n - \delta_k \mathcal L(k)] \|^2 \bE \| \tilde y(k) \|^2 \\
	%		+ \bE \| w(k)\|^2 \\ 
	& +  2 \bE \| \delta_k \nabla_{y} \mathcal P(k+1|k) \|^2 .
\end{aligned}
\end{equation}
From Theorem~\ref{thm: 1}, the estimation errors are bounded within
\begin{equation}\label{upper bound of estimation error}
\bE \| \tilde{ \theta}_{i}(k) \|^2  \leq	\max \left\{\| \tilde{\theta}_{i}(0) \|^2 , \frac{\eta_i^2 L^2 \varrho^2}{1- \max_{j\in\{1,\dots, k-1\}} \rho(A_i(j))}  \right\}.
\end{equation}
As a result, $0\leq \bE \| \delta_k \nabla_{y} \mathcal P(k+1|k) \|^2 \leq  \mu $ is upper bounded, since it is a measure of covariance of the bounded estimators. Consequently, 
%by further noting $\bE \| w(k)\|^2 \leq \rho^2 $, 
we have 
\begin{equation}\label{eqn: expeactation of Euclidean norm dynamics 3}\begin{aligned}
	\bE \|\tilde y(k+1)\|^2 \leq  & 2\| [I_n - \delta_k \mathcal L(k)] \|^2 \bE \| \tilde y(k) \|^2   + \mu.
\end{aligned}
\end{equation}
If there exists a step size $\delta_k$ such that $0< 2\| [I_n - \delta_k \mathcal L(k)] \|^2 <1$, then the expected mean square of the tracking error is convergent.
Recursively iterating \eqref{eqn: expeactation of Euclidean norm dynamics 3} gives 
\begin{equation}
\bE \|\tilde y(k+1)\|^2 \leq  \bar \alpha^k \bE \|\tilde y(0)\|^2 + \sum_{j=0}^{k-1} \bar \alpha^j   \mu
\end{equation}
where $ \bar \alpha := \max_{j\in \{1,\dots, k\}} \alpha_j $ with $0< \alpha_{k} :=	2\| [I_n - \delta_j \mathcal L(j)] \|^2<1$. Since $	\lim_{k\rightarrow \infty} \bar \alpha^k \bE \|\tilde y(0)\|^2 \rightarrow 0$, we have 
\begin{equation}\label{eqn: position MSE}\begin{aligned}
	\lim_{k\rightarrow \infty}  \bE \|y(k) - r^* \|^2 \leq 
	\frac{\mu}{1- \bar \alpha}. 
\end{aligned}
\end{equation}
This completes the proof. 
\end{proof}

\begin{remark}
In general, traditional adaptive control can be regarded as passive learning \cite{mesbah2018stochastic, guay2003adaptive, chai2021dual} where parameter estimators are updated by accidentally collected data samples. For example, MPC in autonomous search is targeted at navigating the agent to the source position, whereas during this pure exploitation process the estimators are updated passively by accidentally collected concentration measurements from the environment \cite{Chen2021DCEE,Li2022ECC}. 
Recently, there are a wide range of engineering problems involved in balancing between exploration and exploitation, e.g., machine learning, control and decision-making in uncertain environment \cite{bar1974dual, tse1973actively, tariverdi2023reinforcement, ghavamzadeh2015bayesian}. In control society, related works are usually focused on stochastic model predictive control with active learning \cite{mesbah2018stochastic}.  A similar concept is referred to as active reinforcement learning in artificial intelligence \cite{ghavamzadeh2015bayesian, jeong2019learning}. Nevertheless, there is a critical distinction between previous works and the proposed DCEE framework for auto-optimisation control. 
{\color{black}
In existing dual control formulation, the probing effect is introduced to learn the \emph{system} states or parameters (e.g. MPC with active learning \cite{mesbah2016stochastic} and active adaptive control \cite{bugeja2009dual, Alpcan2015information}), while in our formulation the probing effect is used to actively explore the operational \emph{environment}. We believe that future autonomous control should be able to deal with not only system uncertainty but also environment uncertainty \cite{Chen2021DCEE, Antsaklis2020autonomy}.}
\end{remark}

\section{DCEE for Linear Systems}\label{sec: DCEE for Linear Systems}
In this section, we deal with general linear systems.
As the environment estimators are designed by information measurements, the parameter adaptation algorithm in \eqref{eqn: environment learning} can be used and Theorem~\ref{thm: 1} remains valid. Now, we design a dual controller that regulates the system output $y(k)$ to minimise the reformulated objective function defined in \eqref{eqn: dual derivation 3}.

The dual controller is proposed as 
\begin{equation}\begin{aligned}\label{eqn: dual controller linear system}
& u(k) = -Kx(k) +(G+ K\Psi) \xi (k) 
\end{aligned}
\end{equation}
where the reference $\xi (k) $ is generated by 
\begin{equation}\begin{aligned}\label{eqn: optimal reference generator}
& \xi(k) = \xi(k-1)  + \psi (k-1) \\
& \psi (k) = -  \delta_k \big[   \nabla_{\xi} \mathcal C(k+1|k) +  \nabla_{\xi} \mathcal P(k+1|k)   \big] 
\end{aligned}
\end{equation}
where $ G $ and $\Psi$ are gain matrices obtained by solving 
\begin{equation}\begin{aligned} \label{eqn: linear matrix equation}
& (A-I)\Psi + BG= 0 \\
& C\Psi - I = 0 .
\end{aligned}
\end{equation}
and $ K$ is chosen such that $A-BK$ is Schur stable as $(A,B)$ is controllable. 
Note that $\psi(k)$ is exactly the dual gradient term used in the integrator dynamics in Section~\ref{sec: DCEE for Integrator}. For linear systems, the control input $ u(k)$ not only needs to have dual effects for exploration and exploitation but additionally requires control effort to stabilise the system dynamics as in \eqref{eqn: dual controller linear system}.

\begin{assumption}\label{asm: rank}
The pair $(A,B)$ is controllable,
and 
\begin{equation}\label{eqn: rank condition}
	\operatorname{rank}\left[\begin{array}{cc}
		A-I & B\\
		C & 0 
	\end{array}\right]=n +q .
\end{equation}
\end{assumption}

{\color{black}
\begin{remark}
The dual control design in \eqref{eqn: dual controller linear system}-\eqref{eqn: linear matrix equation} is partly inspired by conventional internal model approaches \cite{huang2004nonlinear}.  The solvability of \eqref{eqn: linear matrix equation} is guaranteed  by \eqref{eqn: rank condition}, which is widely known as regulation equations \cite{huang2004nonlinear}. The existence of $\Psi$ ensures the existence of optimal state $x^* = \Psi r^*$ such that $Cx^* = r^*$.
\end{remark}
}

Define state transformations $x_s(k) = \Psi \xi(k) $, $u_s(k) = G \xi(k)$. Let $\bar x(k) = x(k)-x_s(k)$ and $\bar u(k) = u(k)-u_s(k)$. Applying the transformation to the system dynamics \eqref{system dyanmics} leads to 
\begin{equation}\label{eqn: transformed dynamics}\begin{aligned}
\bar x(k+1) & =  x(k+1) -x_s(k+1) \\
& = Ax(k)+Bu(k) -\Psi (\xi(k)+\psi(k)) \\ 
& = A\bar x(k) +B\bar u(k) -\Psi \psi(k) \\
e(k) & = C\bar x(k) 
\end{aligned}
\end{equation}
where \eqref{eqn: linear matrix equation} has been used to derive above dynamics. Applying the control input \eqref{eqn: dual controller linear system}, we have the closed loop dynamics 
\begin{equation}\label{eqn: closed-loop error dynamics}\begin{aligned}
\bar x(k+1)  & =  (A-BK)\bar x(k) - \Psi \psi(k) \\
e(k) & = C\bar x(k). 
\end{aligned}
\end{equation}
The following lemma can be regarded as input-to-output stability of the transformed dynamics \eqref{eqn: closed-loop error dynamics} by viewing $\psi(k)$ and $e(k) $ as the input and output, respectively.

\begin{lemma}\label{lem: input to output stability}
Let Assumptions \ref{asm: PE}--\ref{asm: rank} hold. Suppose the conditions specified in Theorems~\ref{thm: 1}--\ref{thm: 2} hold. If the gain matrices $G$ and $\Psi$ are designed according to \eqref{eqn: linear matrix equation} and $K$ is chosen such that $(A-BK)$ is Schur stable, then 
\begin{equation}\label{eqn: input to output stability}
\limsup_{k\rightarrow \infty} \| e(k) \| \leq  \frac{\|C\|\|\Psi\|}{1-\|A-BK\|}   \limsup_{k\rightarrow \infty} \|\psi(k)\|.
\end{equation}
\end{lemma}
\begin{proof}
Putting \eqref{eqn: linear matrix equation} into a matrix form leads to 
\begin{equation}\label{eqn: matrix equation}
\left[\begin{array}{cc}
	A-I & B\\
	C & 0
\end{array} \right]   \left[\begin{array}{c}
	\Psi  \\
	G
\end{array} \right]=  \left[\begin{array}{c}
	0 \\
	I
\end{array} \right]
\end{equation}
of which the solvability is guaranteed under \eqref{eqn: rank condition} in Assumption~\ref{asm: rank} by transforming the matrix equation \eqref{eqn: matrix equation} to standard linear algebraic equations. 
For notational convenience, we denote $A_c = A-BK$ and $B_c = -\Psi$. Then, we have 
\begin{equation}\label{eqn: closed-loop error dynamics 2}\begin{aligned}
	\bar x(k+1)  & =  A_c\bar x(k) +B_c \psi(k) .
\end{aligned}
\end{equation}
Recursively iterating \eqref{eqn: closed-loop error dynamics 2} results in 
\begin{equation}\label{eqn: denition of time index}
\bar x(k) = A_c^k\bar x(0) +\sum_{j=0}^{k-1} A_c^{k-j-1}B_c \psi(j).
\end{equation}
Hence, we have
\begin{equation}\label{eqn: error equation 1}
e(k) = C\bar x(k) =CA_c^k\bar x(0) + C \sum_{j=0}^{k-1} A_c^{k-j-1} B_c \psi(j).
\end{equation}
% where $C\Psi - I =0 $ has been used. 
Because $A_c $ is Schur, we have $\lim_{k\rightarrow \infty} CA_c^k\bar x(0) = 0 $. 

The convergence of reference generator \eqref{eqn: optimal reference generator} has been established in Theorem~\ref{thm: 2}, and thereby $\psi(k)$, i.e., the gradient of the dual controller, is bounded and converges to zero as $k\rightarrow \infty$. Denoting $\varpi := \limsup_{k\rightarrow \infty} \| \psi(k)\| $, it can be obtained that, for any small constant $\epsilon>0$, there exists a positive time index $\zeta>0$ such that 
\begin{equation}\label{eqn: bounds of gradient term}
\| \psi(k)\| < \varpi+\epsilon, \  \forall k>\zeta. 
\end{equation}

Now, the second term in \eqref{eqn: error equation 1} can be separated into two parts, written as 
\begin{equation}\label{eqn: second term separation}\begin{aligned}
C\sum_{j=0}^{k-1} A_c^{k-j-1}B_c \psi(j) = & C\sum_{j=0}^{\zeta} A_c^{k-j-1} B_c\psi(j)\\ & +C\sum_{j=\zeta+1}^{k-1} A_c^{k-j-1}B_c \psi(j).
\end{aligned}
\end{equation}
Taking the Euclidean norm of \eqref{eqn: second term separation} and invoking \eqref{eqn: bounds of gradient term}, we obtain 
\begin{equation}\label{eqn: bounded second term}\begin{aligned}
	&\bigg\|C\sum_{j=0}^{k-1} A_c^{k-j-1}B_c \psi(j)\bigg\| = \|C\|\|B_c\| \bigg\| \sum_{j=0}^{\zeta} A_c^{k-j-1} \psi(j) \\ & \quad \quad  +\sum_{j=\zeta+1}^{k-1} A_c^{k-j-1} \psi(j) \bigg\| \\
	 \leq & \big\| A_c^{k-\zeta-1} \big \|\|C\|\|B_c\| \bigg\| \sum_{j=0}^{\zeta} A_c^{\zeta-j} \psi(j) \bigg \| \\ & +  (\varpi +\epsilon) \|C\|\|B_c\| \bigg\| \sum_{j=\zeta+1}^{k-1} A_c^{k-j-1}  \bigg\| . 
\end{aligned}
\end{equation}
Since $A_c$ is Schur stable (i.e., eigenvalues of $A_c$ are of absolute value less than one), we have 
\begin{equation}
\lim_{k\rightarrow \infty} \big\| A_c^{k-\zeta-1} \big \|  = 0 .
\end{equation}
For the sum of a geometric series, we have 
\begin{equation}
\sum_{j=\zeta+1}^{k-1}\|A_c\|^{k-1-j}=\frac{1-\|A_c\|^{k-\zeta}}{1-\|A_c\|}<\frac{1}{1-\|A_c\|}.
\end{equation}
Therefore, combining \eqref{eqn: error equation 1} and \eqref{eqn: bounded second term} leads to 
\begin{equation}\label{eqn: bound of e}
\limsup_{k \rightarrow \infty} \|e(k) \| \leq \frac{\|C\|\|B_c\|}{1-\|A_c\|}\left(\varpi +\epsilon\right).
\end{equation}
As $\epsilon $ can be set arbitrarily small, it follows from \eqref{eqn: bound of e} that 
\begin{equation}\label{eqn: input to output stability 2}
\limsup_{k\rightarrow \infty} \| e(k) \| \leq \frac{\|C\|\|B_c\|}{1-\|A_c\|}   \limsup_{k\rightarrow \infty} \|\psi(k)\|. 
\end{equation}
%with $\varsigma = \frac{1}{1-\|A_c\|} $. 
This completes the proof.
\end{proof}

Now, combining the results in Theorems~\ref{thm: 1}--\ref{thm: 2} and Lemma~\ref{lem: input to output stability}, we are ready to establish the convergence of the auto-optimisation control for linear systems. 

\begin{theorem}
Let Assumptions~\ref{asm: PE}--\ref{asm: rank} hold. Suppose the conditions specified in Theorems~\ref{thm: 1}--\ref{thm: 2} and Lemma~\ref{lem: input to output stability} hold. The output $y(k)$ of the linear system \eqref{system dyanmics} converges to the neighbourhood of the optimum $r^*$, using control input \eqref{eqn: dual controller linear system} together with reference generator \eqref{eqn: optimal reference generator}. 
%Moreover, in the absence of measurement noises, $y(k)$ converges to the true optimal solution $r^*$. 
\end{theorem}
\begin{proof}
Denoting $\tilde x(k) = x(k)-\Psi r^*$,  we have 
\begin{equation}\begin{aligned} 
	\tilde x(k+1) & = Ax(k) + B [-Kx(k) +(G+ K\Psi) \xi (k) ] - \Psi r^* \\
	& =  (A-BK) \tilde x(k) + B(G+K\Psi) (\xi(k) - r^*)
\end{aligned}
\end{equation}
It follows from Theorems~\ref{thm: 1}--\ref{thm: 2} that $\xi(k) $ converges to the neighbourhood of $r^*$ with bounded error. Thus, the result can be concluded by treating $B(G+K\Psi) (\xi(k) - r^*)$ as $\psi(k)$ in Lemma~\ref{lem: input to output stability}. 
\end{proof}

\begin{remark}
The auto-optimisation control in this paper is similar to the classic formulation of reinforcement learning in the sense that both of them are targeted to operate a system in an unknown and uncertain environment. There are two bottlenecks in widely applying reinforcement learning, particularly deep RL: one is a large number of trials are required to achieve a satisfactory performance (big data) and the other is its performance could significantly degrade if the real operational environment is different from the training environment (poor adaptiveness) \cite{Chen2022perspective}. DCEE establishes a new control framework to provide a promising and complementary method to reinforcement learning in control and robotics society, which can deal with uncertain environment without repetitive training. In fact, active learning for exploration and exploitation in machine intelligence can find strong evidence in human intelligence, which is supported by the biological principles in functional integration in the human brain and neuronal interactions (known as free-energy principle and active inference in neuroscience \cite{friston2010free}). Interested readers are referred to \cite{Chen2022perspective} for detailed discussions. 
\end{remark}

\section{Numerical Example}\label{sec: Numerical Example}
In this section, we verify the effectiveness of the proposed algorithm using a dedicate numerical example. Consider a linear system \eqref{system dyanmics} with 
\begin{equation}
A = \left[\begin{array}{cc}
0  & 1 \\
2 & 1
\end{array} \right] , \ 
B = \left[\begin{array}{c}
1 \\
1
\end{array} \right] , \  C = \left[\begin{array}{cc}
0  &	1
\end{array}\right] .
\end{equation}
%with an additive white noises $w(k) \sim \mathcal N (0, 0.1) $.
The reward function is given by 
\begin{equation}
J(\theta^*, y) = 2y - \theta^* y^2 = \left[\begin{array}{cc}
2y  & -y^2 
\end{array} \right] \left[\begin{array}{c}
1  \\ \theta^*
\end{array} \right] 
%	[2y, -y^2] [1, \theta^*]^{\tp}
\end{equation}
where $\theta^* $ is affected by the unknown environment. The true value is $\theta^*=1$ but unavailable \emph{a priori}. The optimal operational condition $r^*$ is determined by $\theta^*$, i.e., $r^* = l(\theta^*) = 1/\theta^*=1$. 

We assume the measurements are subject to Gaussian noise $v(k) \sim \mathcal N(0, 2) $, which implies that the observations from environment are $ J(k) = J(\theta^*, y(k)) +v(k) $. Decision-making under uncertain environment with noisy measurements is of significant importance to promote the system intelligence. 
In order to explore  the uncertain environment, the first step is to quantify the level of uncertainty. An ensemble based multi-estimator approach has been developed in previous sections. Now,
the size of the estimator ensemble is chosen as $ N=100 $, and each of them is randomly initialised according to a uniform distribution between $-10$ and $10$, i.e., $ \theta_i(0) \sim U(-10,10),  \forall i =1,2,\dots, 100$. 
{\color{black}
Choosing the number of the ensemble necessitates balancing computational efficiency with stochastic accuracy. Depending on the specific problem and available resources, it is common practice to experiment with various configurations until a satisfactory balance is achieved. Indeed, those findings align with recent trends observed in machine learning community using ensemble aggregating approaches where a handful of estimators are enough to generate promising results \cite{chua2018deep, Lakshminarayanan2017simple}.}
%The step sizes are set as $\eta_i=0.005 $ and $\delta_k = 0.5$. 
The system is controllable and regulation condition in~\eqref{eqn: rank condition} is satisfied such that the gain matrices can be obtained as $ \Psi = [\frac 13, 1]^{\tp}$ and $ G= - \frac 23 $. The gain matrix $K = [-1, 1.1] $ is chosen by placing the poles of $(A-BK) $  at $[0.4;0.5]$.

\begin{figure*}
\centering
\begin{subfigure}[b]{0.45\textwidth}
\centering
\includegraphics[width=1\textwidth]{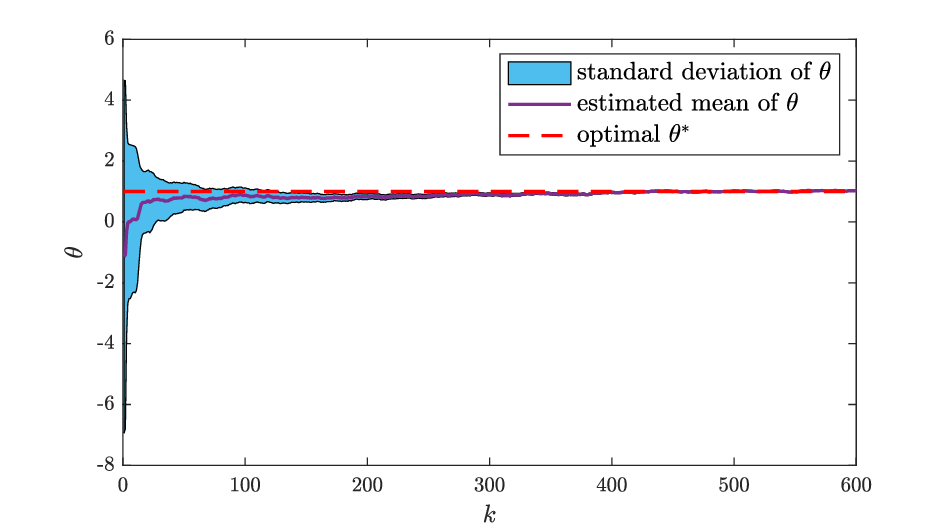}
\caption{Mean and standard deviation of estimated $\theta(k)$.}
\label{fig: DCEE_Theta}
\end{subfigure}
\begin{subfigure}[b]{0.45\textwidth}
\centering
\includegraphics[width=1\textwidth]{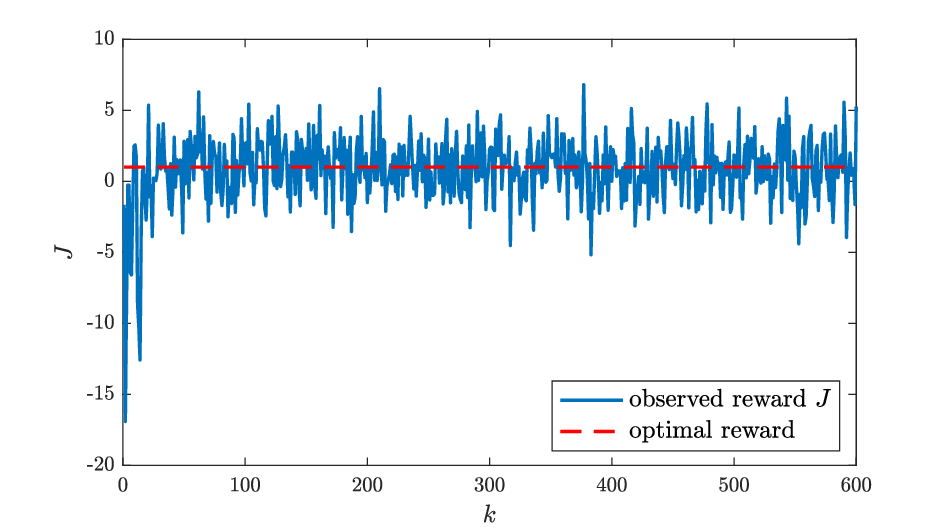}
\caption{Observed reward $J(k)$ with measurement noises $v(t)$.}
\label{fig: DCEE_Reward}
\end{subfigure}

\begin{subfigure}[b]{0.45\textwidth}
\centering
\includegraphics[width=1\textwidth]{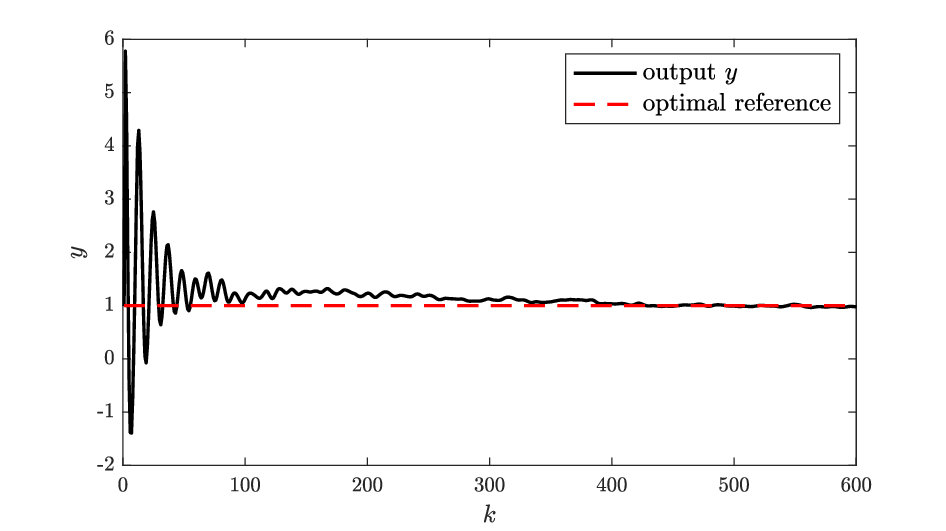}
\caption{System output $y(k)$.}
\label{fig: DCEE_y}
\end{subfigure}
\begin{subfigure}[b]{0.45\textwidth}
\centering
\includegraphics[width=1\textwidth]{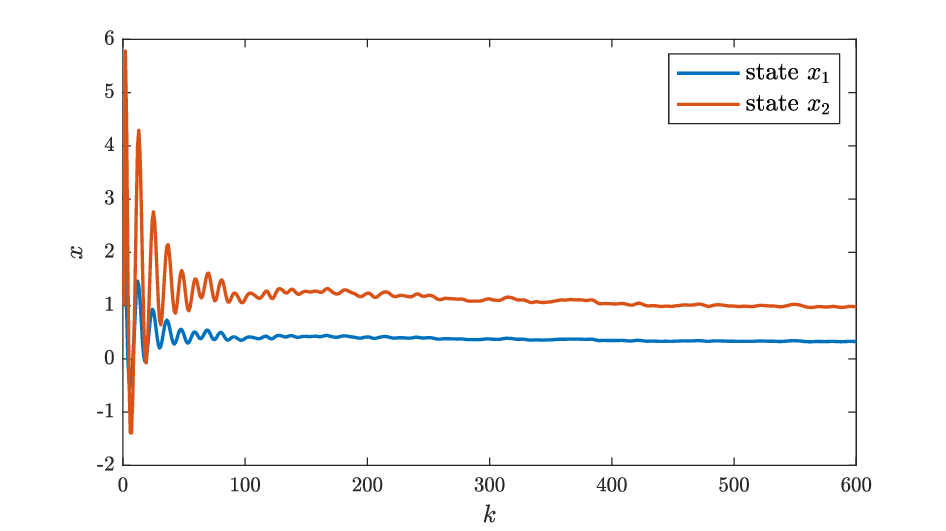}
\caption{System state $x(k)$.}
\label{fig: DCEE_X}
\end{subfigure}
\caption{Simulation results with environment uncertainties using active learning based DCEE.}
\label{fig: active learning}
\end{figure*}

\begin{figure*}
\centering
\begin{subfigure}[b]{0.45\textwidth}
\centering
\includegraphics[width=1\linewidth]{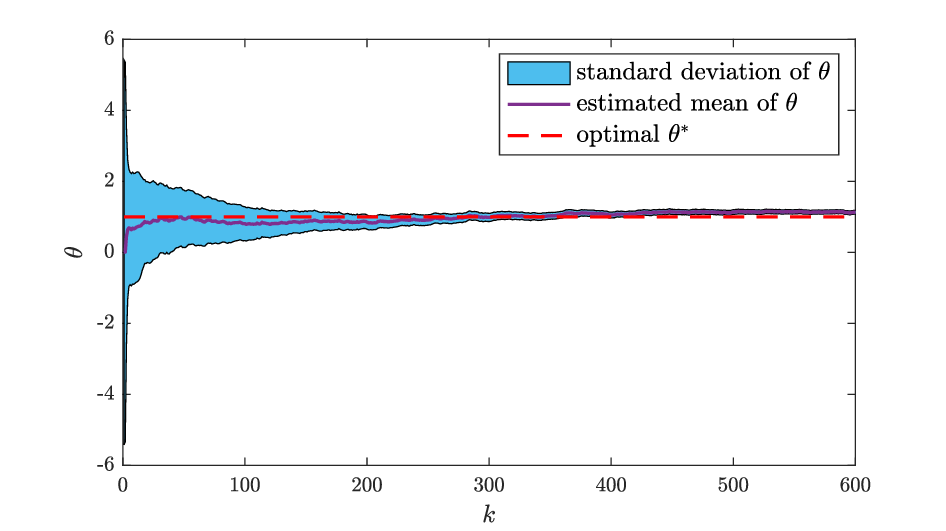}
\caption{Mean and standard deviation of estimated $\theta(k)$.}
\label{fig: estimation_theta_noise_free}
\end{subfigure}
\begin{subfigure}[b]{0.45\textwidth}
\centering
\includegraphics[width=1\linewidth]{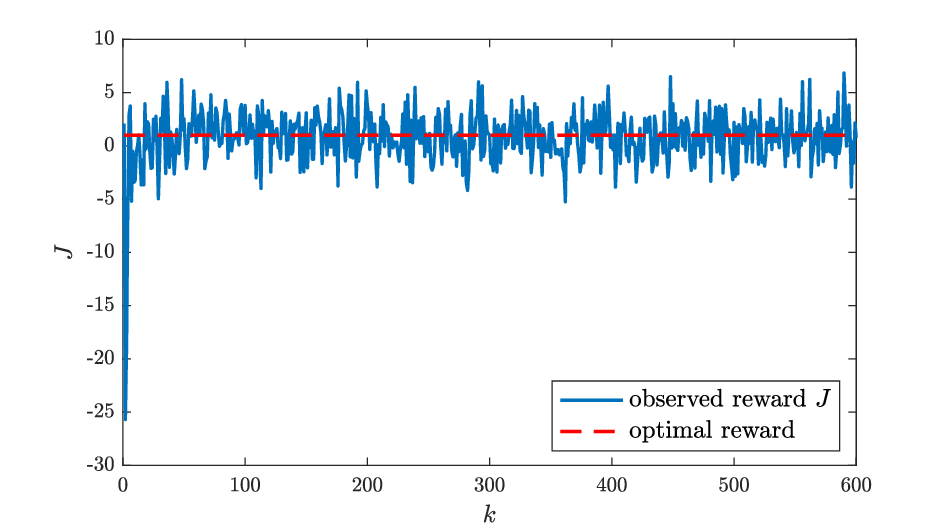}
\caption{Observed reward $J(k)$ with measurement noises $v(t)$.}
\label{fig: observed_reward_noise_free}
\end{subfigure}

\begin{subfigure}[b]{0.45\textwidth}
\centering
\includegraphics[width=1\linewidth]{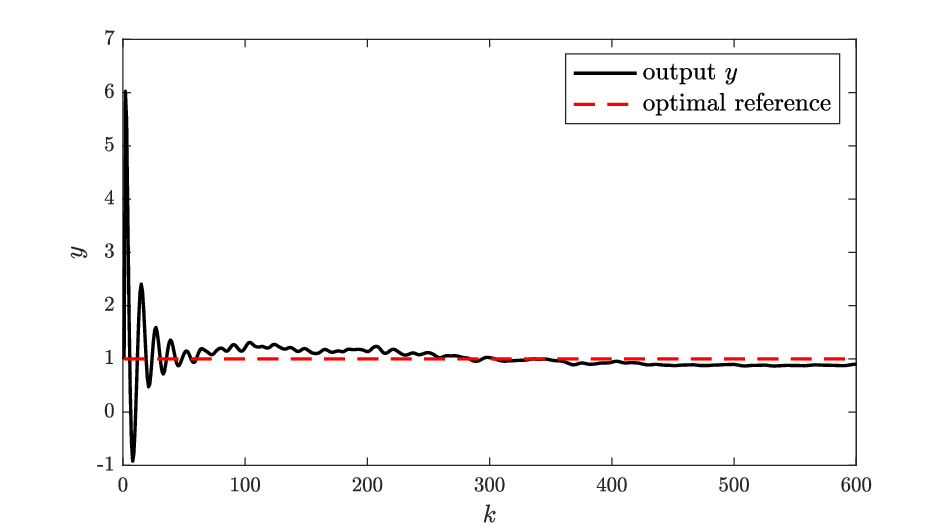}
\caption{System output $y(k)$.}
\label{fig: y_noise_free}
\end{subfigure}
\begin{subfigure}[b]{0.45\textwidth}
\centering
\includegraphics[width=1\linewidth]{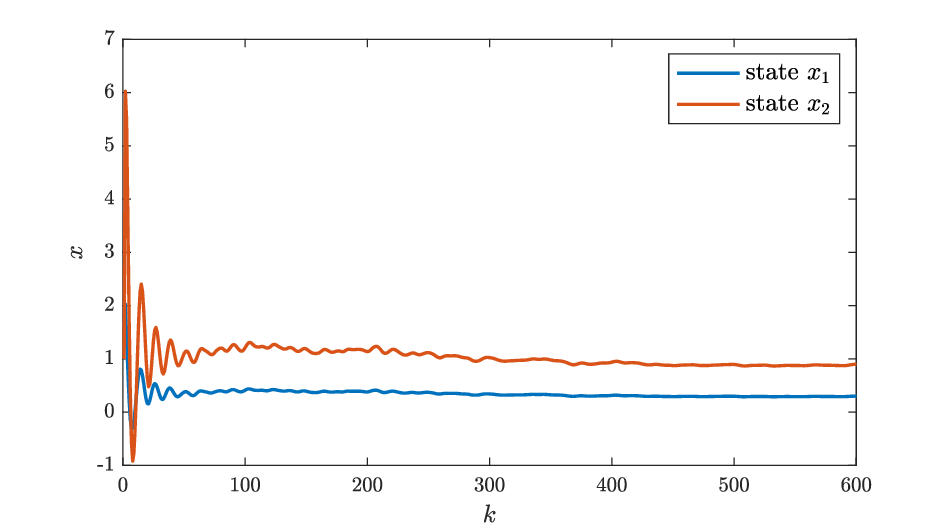}
\caption{System state $x(k)$.}
\label{fig: state_noise_free}
\end{subfigure}
\caption{Simulation results with environment uncertainties using passive learning based auto-optimisation approach.}
\label{fig: passive learning}
\end{figure*}

{\color{black}
Fig.~\ref{fig: DCEE_Theta} shows the estimated environmental parameters. Initially, the mean and standard deviation of the ensemble $\{\theta_{i}, i=1,\dots, 100\}$ are $-1.137$ and $5.788$, respectively, randomly initialised using a uniform distribution. The mean of the estimators converges to the true environment parameter $\theta^*=1$, and the standard deviation among the estimators shrinks quickly, indicating that the estimation uncertainty reduces (quantified by the variance among the estimators in the ensemble). Despite increasing the iteration $k$ significantly, the estimated parameters remain fluctuating within a  small neighbourhood of the true value due to the presence of noisy measurements. Fig.~\ref{fig: DCEE_Reward} displays the observed rewards from the environment. Even though we have imposed quite significant noises to the measurements, the performance of the estimators is fairly satisfactory, which manifests the ensemble based active learning provides superior robustness against noises. 

Implementing the dual control in \eqref{eqn: dual controller linear system} not only contributes to enhanced parameter adaptation performance but also drives the system output to the optimal operational condition, as shown in Fig.~\ref{fig: DCEE_y}. 
The system output approaches the optimal operational point $r^* = 1$ as shown in Fig. \ref{fig: DCEE_y}, and the system states are displayed in Fig. \ref{fig: DCEE_X}. It can be verified that $x^* = \Psi r^* =  [\frac 13, 1]^{\tp}$.
The tracking error is determined by the estimation error. In this process, there is no need to tune the weights of exploration and exploitation. As a principled approach, the dual controller in \eqref{eqn: dual controller linear system} is derived from a physically meaningful objective function, which naturally embeds balanced dual effects for active environment learning and optimality tracking. 

To demonstrate the impact of active learning based exploration mechanism, we have implemented certainty equivalence based auto-optimisation approach using passive learning. For fare comparisons, we have kept all parameters the same but removed the exploration effect from the control input. The obtained results are depicted in Fig. \ref{fig: passive learning}. A key observation to highlight is that incorporating active learning into the DCEE framework significantly enhances rapid environmental acquisition and eliminates steady-state discrepancies. These improvements are notable when compared to passive auto-optimisation control approaches, which often suffer from suboptimal learning performance caused by inaccurate estimation result. Furthermore, our findings challenge the effectiveness of the certainty equivalence principle commonly used in classic adaptive control, particularly in complex and uncertain environments. This validation underscores the need for more advanced control strategies like DCEE in handling uncertainty.}

\section{Application for MPPT}\label{sec: MPPT}
DCEE was originally developed to solve autonomous search problem in \cite{Chen2021DCEE}, which demonstrates outstanding performance compared with other existing approaches. In this section, we take the optimal control for photovoltaic (PV) systems as an example to illustrate that DCEE can be implemented to solve a much wider class of auto-optimisation control problems in real-world applications. 
Extracting maximum power is a long-lasting pursuit in operating PV systems. Despite significant research efforts made over the past few decades \cite{Bhatnagar2013maximum, Reisi2013classification, Esram2007comparison}, the energy conversion efficiency of PV systems remains very poor due to high environment uncertainties in temperature, irradiance level, partial shading and other atmospheric conditions. The primary goal in PV operation is simply to extract solar energy as much as possible despite changing operational environment, termed as maximum power point tracking (MPPT). There have been a wide variety of methods targeting to solve this problem, which can be roughly classified into three categories: offline methods, online methods, and other methods.  Detailed comparisons and classifications can be found in comprehensive survey papers, e.g., \cite{Bhatnagar2013maximum, Reisi2013classification}. 

In this section, the proposed DCEE is implemented as an alternative approach to achieve MPPT, and two representative approaches, hill climbing method (HC) and incremental conductance method (IC), are deployed for comparison.  It is worth noting that all the three algorithms can be classified as online methods. It has been widely perceived that online methods usually outperform offline counterparts in terms of conversion efficiency due to their inherent adaptiveness to changing environment.  According to the curve-fitting based MPPT \cite{Bhatnagar2013maximum}, the power and voltage ($P$-$V$) characteristics can be modelled by 
\begin{equation}\label{eqn: P-V model}
P = \phi^{\tp}(V) \theta 
\end{equation}
where $ \phi(V) $ is the polynomial regressor $[1, V, V^2,\dots, V^n ]^{\tp}$ and $\theta \in \bR^{n+1}$ is the polynomial coefficient. To solve the maximum problem of \eqref{eqn: P-V model}, we need to estimate the unknown parameters $\theta$ and then maximise the power output by regulating the voltage $V$ according to
\begin{equation}
V(k+1) = V(k) + u(k). 
\end{equation}

\begin{figure}
\centering
\includegraphics[width=1\linewidth]{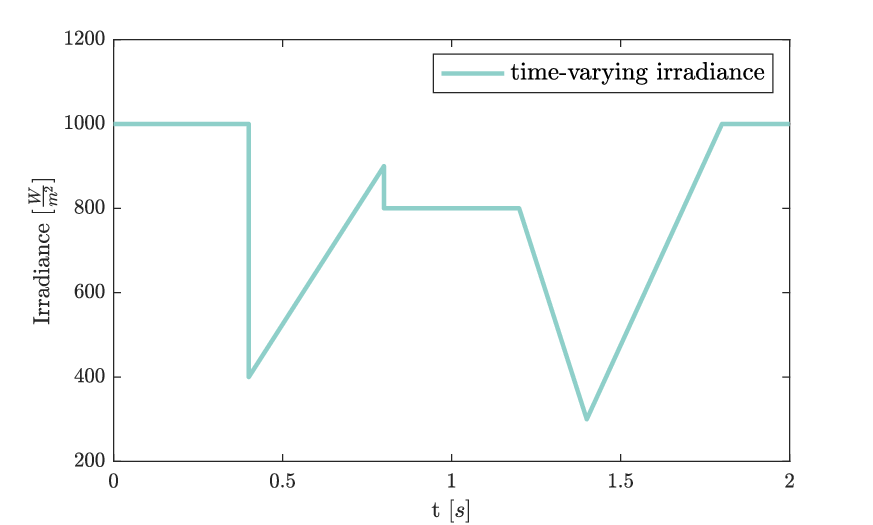}
\caption{Time-varying solar irradiance profile.}
\label{fig: irradiance}
\end{figure}

We use solar panel A10Green Technology model number A10J-S72-175 for this simulation \cite{shams2020maximum}. To mimic the real operational environment of PV systems,  a time-varying solar irradiance profile is stimulated as shown in Fig.~\ref{fig: irradiance}, and the temperature is initially set as 25\textcelsius \ and then jumps to 35\textcelsius \  at $t=1$s. It should be noted that the unknown environment parameter $\theta$ changes as the operational condition varies.
Although the proposed algorithm is theoretically analysed for static parameters identification, the use of constant learning rate  $\eta_i$ renders the adaptation algorithm in \eqref{eqn: environment learning} with the capability of tracking drifting parameters. 

\begin{figure}
\centering
\includegraphics[width=1\linewidth]{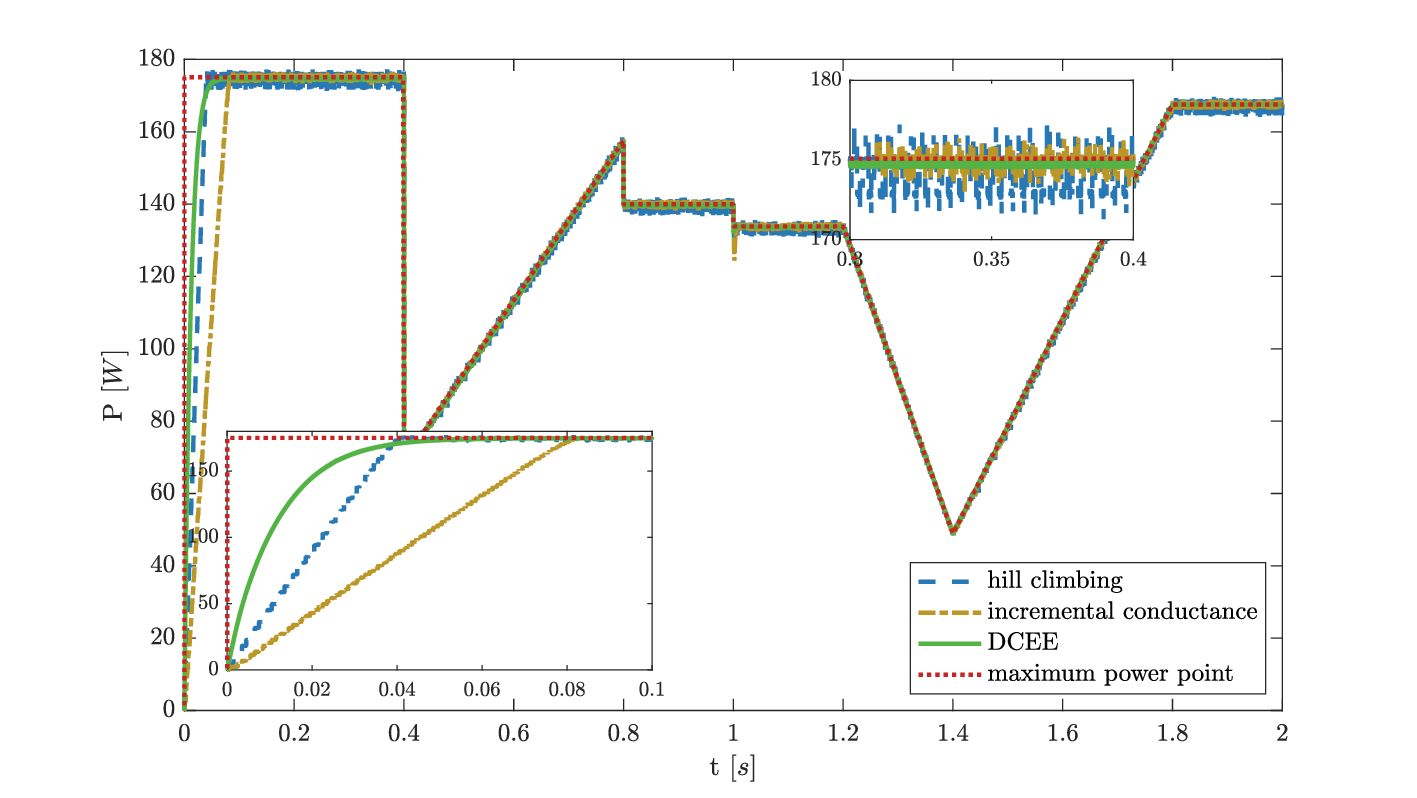}
\caption{Power profile using different algorithms.}
\label{fig: power}
\end{figure}

\begin{figure}
\centering
\includegraphics[width=1\linewidth]{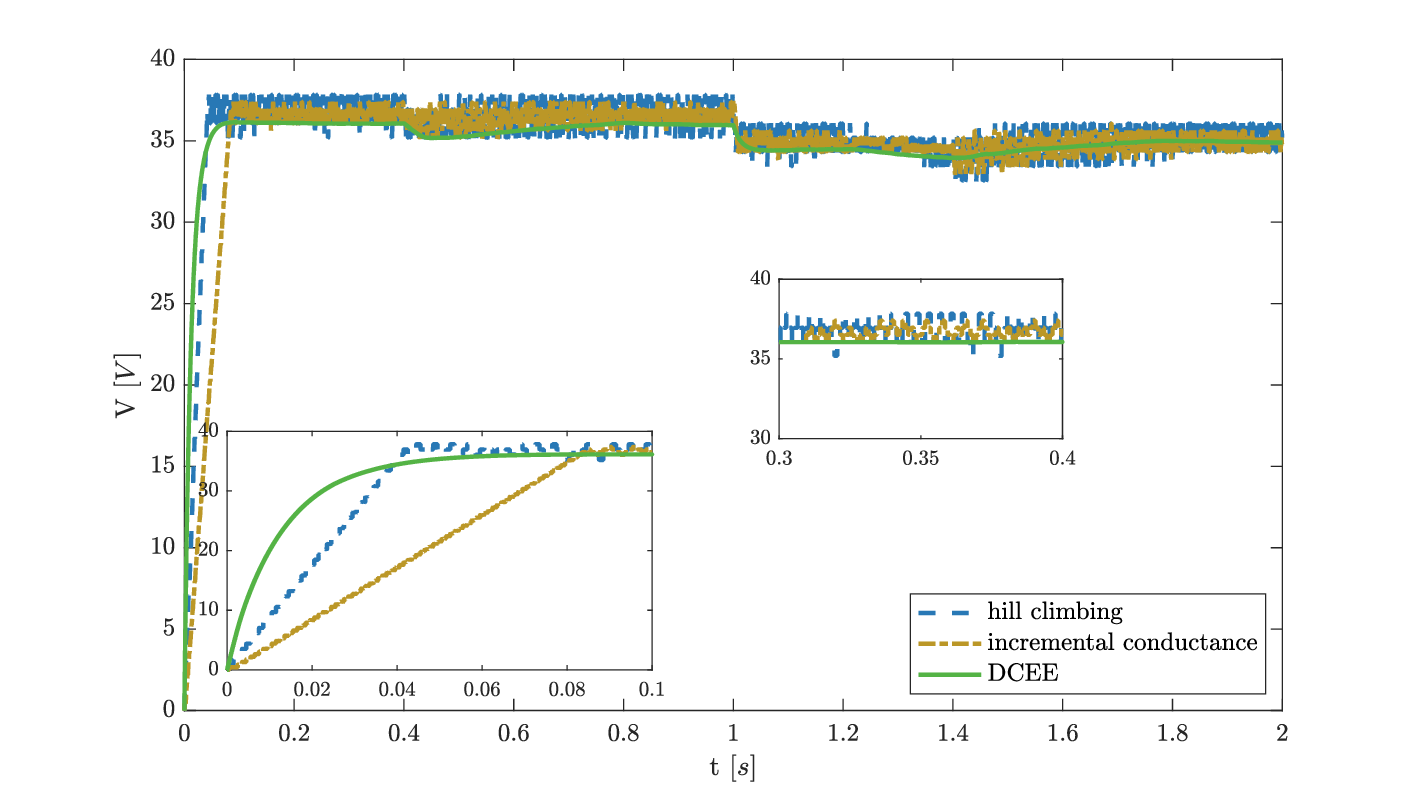}
\caption{Voltage profile using different algorithms.}
\label{fig: voltage}
\end{figure}

\begin{figure}
\centering
\includegraphics[width=1\linewidth]{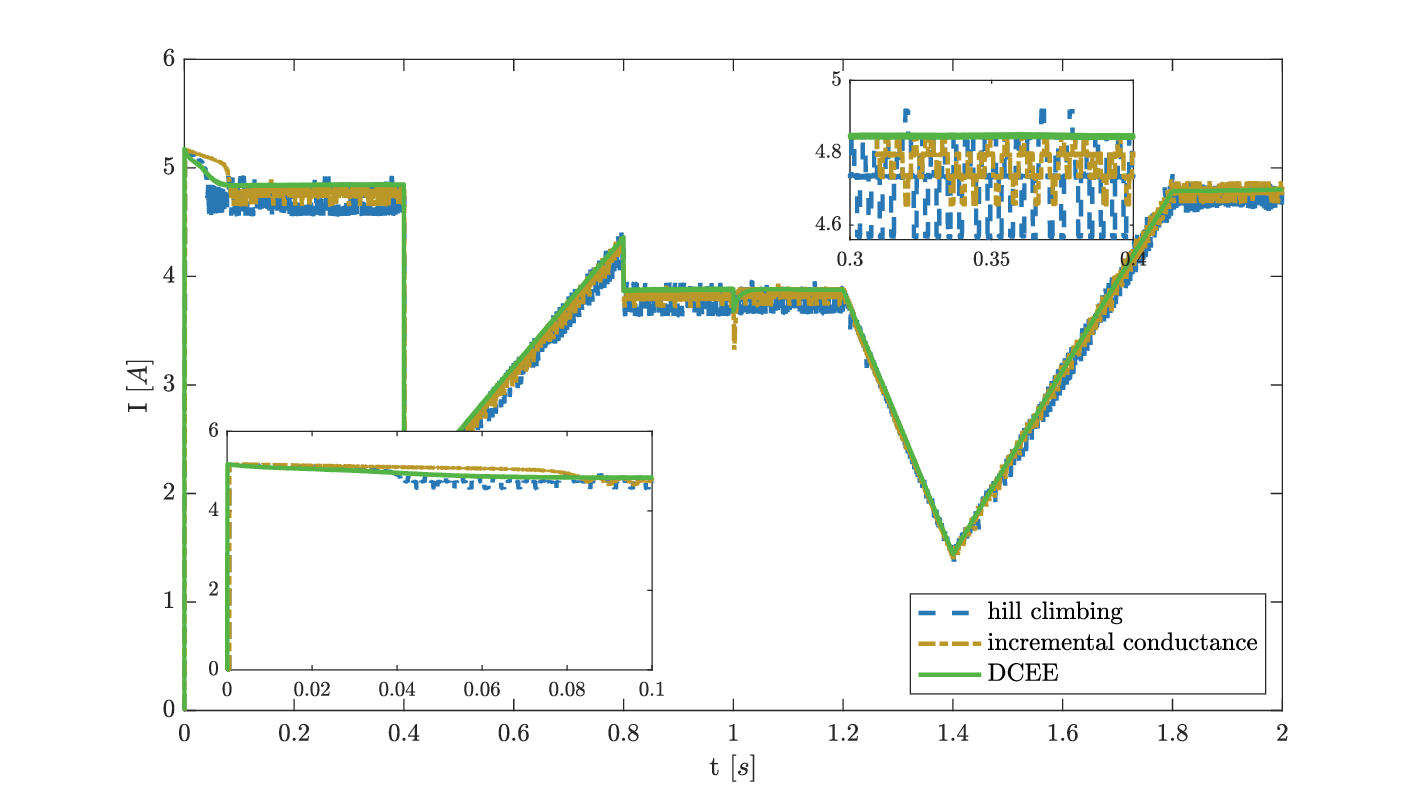}
\caption{Current profile using different algorithms.}
\label{fig: current}
\end{figure}

\begin{figure}
\centering
\includegraphics[width=1\linewidth]{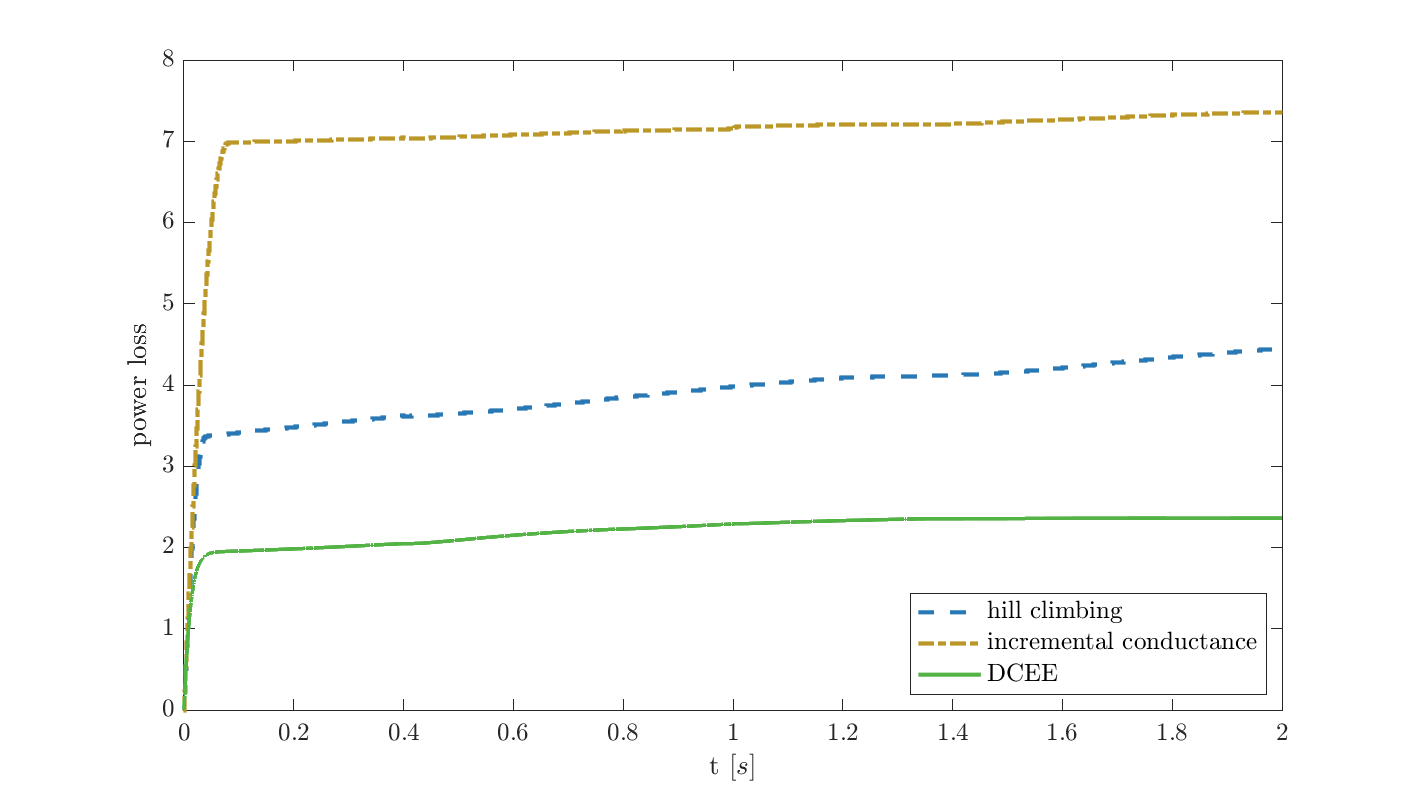}
\caption{Power losses using different algorithms.}
\label{fig: power loss}
\end{figure}

Simulation results using different algorithms (DCEE, HC and IC) are shown in Fig. \ref{fig: power}, \ref{fig: voltage} and \ref{fig: current}. To illustrate more detailed features of different algorithms, enlarged sub-figures are displayed for the time intervals $t\in[0,0.1] $, and $t\in[0.3, 0.4]$.  
The power losses, as displayed in Fig. \ref{fig: power loss}, are calculated by integrating the differences between the maximum power point and real power outputs stimulated using different algorithms.  
Convergence speed,  sensed signals, algorithm complexity and conversion efficiency are four commonly-used criteria to assess the characteristics of MPPT techniques. According to the simulation results, we summarise and compare the features of different approaches in Table~\ref{tab: algorithm features}. Conversion efficiency directly influences the energy extracted from the PV systems, which is ratio between real generated energy and maximum energy (accumulated over the simulation time interval $[0,2]$).  DCEE produces quite high efficiency (99.1\%). Due to the use of perturbed signals in hill climbing method, there are very large voltage and current fluctuations in steady state. This undesirable property not only causes low conversion efficiency but also leads to fast degradation in low level electronic devices. The oscillations are partially solved by incremental conductance method, which measures incremental current and voltage changes to predict the effect of voltage change.

 Different from HC, incremental inductance method is able to maintain at MPP without oscillations when there is no change in operational environment.  From the simulation results using HC and IC, there is a trade-off between transient convergence speed and steady-state oscillations. The steady-state oscillation of IC is reduced at the cost of slow tracking performance, leading to larger power loss with a conversion efficiency $97.2\%$.  It is argued that DCEE as a balanced approach is able to optimally trade-off between exploitation and exploration: when there is large uncertainty in estimated MPP, it will explore quickly to gain information to construct more accurate estimate of MPP; and when there is less change in operational environment, it will maintain at the current belief of MPP without causing large oscillations.  
All three algorithms need to measure voltage and current: DCEE requires voltage and power (calculated by the product of current and voltage) to construct $P$-$V$ curve in \eqref{eqn: P-V model} (i.e., reward-state mapping), while HC and IC use incremental power to decide the direction of voltage regulation. As mature MPPT techniques, both HC and IC are simple to implement using dedicated hardware devices. Since efficient ensemble approximation and gradient based control are developed in this new approach, DCEE is ready to be implemented in real PV platforms without incurring heavy computational load.  

\begin{table*}
\centering
\caption{Features of different MPPT techniques.}
\label{tab: algorithm features}
%	\resizebox{\textwidth}{!}{
\begin{tabular}{@{}llllllll@{}}
\toprule
&		Methods	 & Convergence speed & Sensed variables & Algorithm complexity & Conversion efficiency \\ \midrule
1 & DCEE  & Fast   & Voltage and current  & Medium  &  99.1\%\\ 
2 & Hill climbing		 	& Fast				& Voltage and current				& Simple  & 98.3\%\\	 	
3 & Incremental conductance  & Medium &	Voltage and current	  & Simple & 97.2\%
\\ \bottomrule
\end{tabular}
%		}
\end{table*}

\section{Conclusion}\label{sec: Conclusion}
This paper has proposed a dual control framework for exploration and exploitation, designed to address auto-optimisation control challenges in uncertain environments. The DCEE algorithm is uniquely structured to balance exploration and exploitation, resolving the inherent conflict between parameter identification and the pursuit of optimal operation. We have rigorously demonstrated that our approach ensures convergence and maintains performance, considering both the objective function and the characteristics of environmental noise. The effectiveness of the DCEE framework is further evidenced through a numerical example and its successful application in MPPT, underscoring its substantial potential for diverse real-world scenarios.

\section*{Acknowledgment}
For the purpose of open access, the author(s) has applied a Creative Commons Attribution (CC BY) license to any Accepted Manuscript version arising.

\bibliographystyle{IEEEtran}

\bibliography{AutoOptControl}

\begin{IEEEbiography}[{\includegraphics[width=1in,height=1.25in,clip,keepaspectratio]{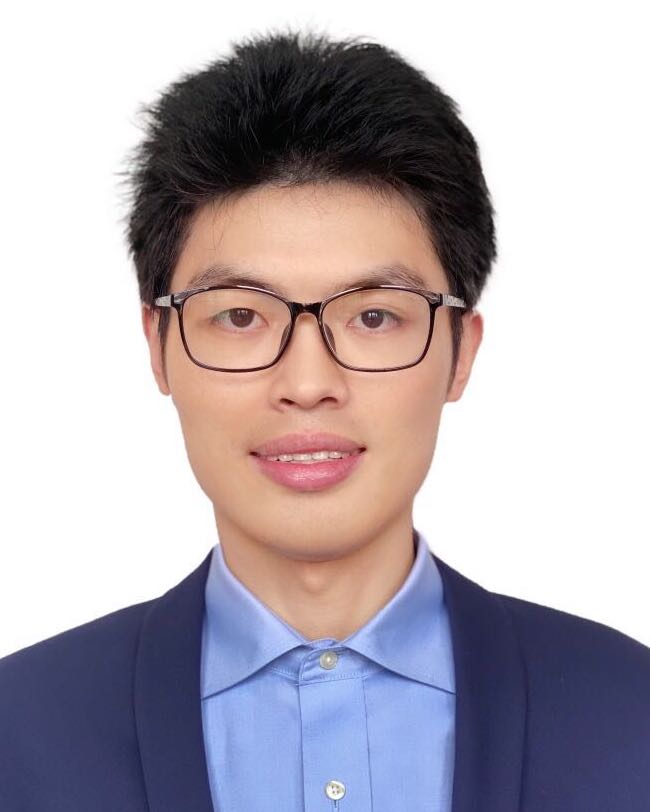}}]
	{Zhongguo Li} (Member, IEEE) received the B.Eng. and Ph.D. degrees in electrical and electronic engineering from the University of Manchester, Manchester, U.K., in 2017 and 2021, respectively.
	
    He is currently a Lecturer in Robotics, Control, Communication and AI at the University of Manchester. Before joining Manchester, he was a Lecturer at University College London and a Research Associate at Loughborough University. His research interests include optimisation and decision-making for advanced control, distributed algorithm development for game and learning in network connected multi-agent systems, and their applications in robotics and autonomous vehicles.
\end{IEEEbiography}

\begin{IEEEbiography}[{\includegraphics[width=1in,height=1.25in,clip,keepaspectratio]{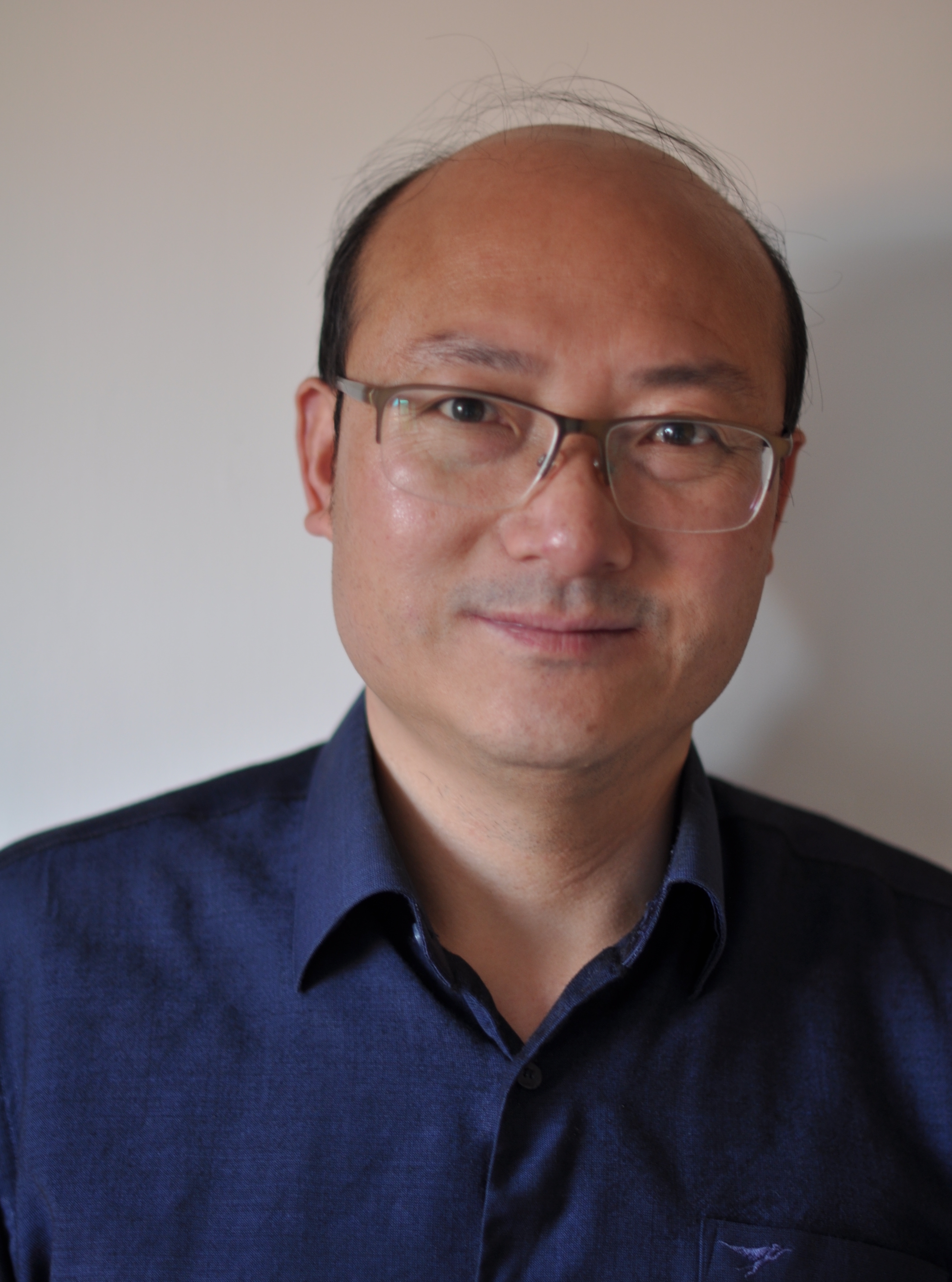}}]
	{Wen-Hua Chen} (Fellow, IEEE) holds Chair in Autonomous Vehicles with the Department of Aeronautical and Automotive Engineering, Loughborough University, U.K. He is the founder and the Head of the Loughborough University Centre of Autonomous Systems. He is interested in control, signal processing and artificial intelligence and their applications in robots, aerospace, and automotive systems. Dr Chen is a Chartered Engineer, and a Fellow of IEEE, the Institution of Mechanical Engineers and the Institution of Engineering and Technology, U.K. He has authored or coauthored near 300 papers and 2 books. Currently he holds the UK Engineering and Physical Sciences Research Council (EPSRC) Established Career Fellowship in developing new control theory for robotics and autonomous systems.

\end{IEEEbiography}

\begin{IEEEbiography}[{\includegraphics[width=1in,height=1.25in,clip,keepaspectratio]{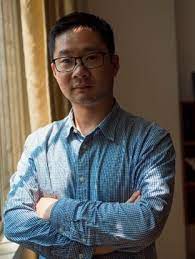}}]
{Jun Yang} (Fellow, IEEE) received the B.Sc. degree in automation from the Department of Automatic Control, Northeastern University, Shenyang, China, in 2006, and the Ph.D. degree in control theory and control engineering from the School of Automation, Southeast University, Nanjing, China, in 2011. 

He joined the Department of Aeronautical and Automotive Engineering at Loughborough University in 2020 as a Senior Lecturer and was promoted to a Reader in 2023. His research interests include disturbance estimation and compensation, and advanced control theory and its application to mechatronic control systems and autonomous systems. 
He serves as Associate Editor or Technical Editor of IEEE Transactions on Industrial Electronics, IEEE-ASME Transactions on Mechatronics, IEEE Open Journal of Industrial Electronics Society, etc. He was the recipient of the EPSRC New Investigator Award. He is a Fellow of IEEE, IET and AAIA.
\end{IEEEbiography}

\begin{IEEEbiography}[{\includegraphics[width=1in,height=1.25in,clip,keepaspectratio]{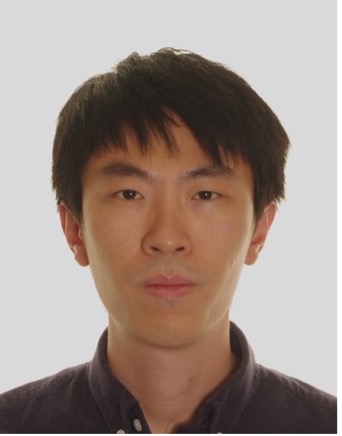}}]
{Yunda Yan} (Member, IEEE) received the B.Sc. degree in automation and the Ph.D. degree in control theory and control engineering from the School of Automation in Southeast University, Nanjing, China, in 2013 and 2019, respectively. 

From 2020 to 2022, he was a Research Associate with the Department of Aeronautical and Automotive Engineering, Loughborough University, U.K.  From 2022 to 2023, he was with the School of Engineering and Sustainable Development, De Montfort University, U.K as a Lecturer in Control Engineering and was later promoted to a Senior Lecturer.  In Sep. 2023, he joined the Department of Computer Science, University College London, U.K, as a Lecturer in Robotics and AI. His current research interest focuses on the safety-critical control design for autonomous systems, especially related with optimisation and learning-based methods. 
\end{IEEEbiography}

\end{document}